\documentclass[sn-aps]{sn-jnl}

\usepackage{hyperref}
\hypersetup{linkcolor=blue,citecolor=red}

\usepackage{natbib}
\usepackage{amssymb,amsmath}

\usepackage{epsfig}
\usepackage{bm}
\usepackage{color}
\usepackage{graphicx}

\newcommand\beq{\begin{equation}}
\newcommand\eeq{\end{equation}}
\newcommand\beqa{\begin{eqnarray}}
\newcommand\eeqa{\end{eqnarray}}
\def\bal#1\eal{\begin{align}#1\end{align}}
\newcommand{\nn}{\nonumber\\}

\newcommand{\bc}{\mathbf{v}}
\newcommand{\bV}{\mathbf{V}}
\newcommand{\ka}{\kappa}
\newcommand{\bt}{\widetilde\beta}
\newcommand{\at}{\widetilde\alpha}

\newcommand{\bk}{\hnn{\sigma}}
\newcommand{\medio}[1]{\left\langle #1\right\rangle}
\newcommand{\al}{\alpha}
\newcommand{\be}{\beta}

\newcommand{\bx}{\mathbf{r}}
\newcommand{\buu}{\mathbf{u}}
\newcommand{\bw}{\boldsymbol{\omega}}
\newcommand{\bW}{\boldsymbol{\Omega}}
\newcommand{\bxi}{\boldsymbol{\Gamma}}

\newcommand{\nuM}{\nu}
\newcommand{\hnn}[1]{\widehat{\boldsymbol #1}}
\newcommand{\Dert}{\mathcal{D}_t}
\newcommand{\bbA}{\boldsymbol{\mathcal{A}}}
\newcommand{\bbB}{\boldsymbol{\mathcal{B}}}
\newcommand{\bbC}{\mathcal{C}}
\newcommand{\bbE}{\mathcal{E}}

\jyear{2024}%

\theoremstyle{thmstyleone}%
\theoremstyle{thmstyletwo}%

\theoremstyle{thmstylethree}%

\begin{document}

\title[Transport coefficients from the inelastic rough Maxwell model]{Exact transport coefficients from the inelastic rough Maxwell model of a granular gas}


\author*[1]{\fnm{Andr\'es} \sur{Santos}}\email{andres@unex.es}
\author[2]{\fnm{Gilberto M.} \sur{Kremer}}\email{kremer@fisica.ufpr.br}

\affil[1]{\orgdiv{Departamento de F\'{\i}sica and Instituto de Computaci\'on Cient\'ifica Avanzada (ICCAEx)}, \orgname{Universidad de Extremadura},
\orgaddress{
\city{Badajoz}, \postcode{E-06006},
\country{Spain}}}

\affil[2]{\orgdiv{Departamento de F\'{\i}sica}, \orgname{Universidade Federal do Paran\'a},
\orgaddress{
\city{Curitiba},
\country{Brazil}}}



\abstract{Granular gases demand models capable of capturing their distinct characteristics. The widely employed inelastic hard-sphere model (IHSM) introduces complexities that are compounded when incorporating realistic features like surface roughness and rotational degrees of freedom, resulting in the more intricate inelastic rough hard-sphere model (IRHSM). This paper focuses on the inelastic rough Maxwell model (IRMM), presenting a more tractable alternative to the IRHSM and enabling exact solutions. Building on the foundation of the inelastic Maxwell model (IMM) applied to granular gases, the IRMM extends the mathematical representation to encompass surface roughness and rotational degrees of freedom. The primary objective is to provide exact expressions for the Navier--Stokes--Fourier transport coefficients within the IRMM, including the shear and bulk viscosities, the thermal and diffusive heat conductivities, and the cooling-rate transport coefficient. In contrast to earlier approximations in the IRHSM, our study unveils inherent couplings, such as shear viscosity to spin viscosity and heat conductivities to counterparts associated with a
torque-vorticity vector. These exact findings provide valuable insights into refining the Sonine approximation applied to the IRHSM, contributing to a deeper understanding of the transport properties in granular gases with realistic features.}




\maketitle

\section{Introduction}\label{sec1}
While it is well known that a hydrodynamic description can be applied to granular gases \cite{D01,BP04,G19}, the derivation of the associated Navier--Stokes--Fourier (NSF) transport coefficients faces important challenges. To begin with, the spatially uniform and isotropic base state is not that of equilibrium (as in the case of molecular gases) but either the so-called homogeneous cooling state (HCS) in the absence of any external driving  or a nonequilibrium steady state if energy is injected to the granular gas by a certain thermostat. In top of that, one usually needs to leave out some realistic properties of the grains and focus on their most basic features by choosing a physically relevant model. In the most widely used model, the grains are assumed to be represented by perfectly smooth spheres characterized by a constant coefficient of normal restitution, $\alpha$ \cite{C90}. Despite that, the derivation of the NSF transport coefficients in this inelastic hard-sphere model (IHSM) requires the use of nontrivial approximations \cite{BDKS98,GDH07,GSM07,G19}. Those approximations become even more problematic if the hard-sphere model is made more realistic by assuming surface roughness, so that the rotational degrees of freedom need to be  included in the description. The resulting inelastic rough hard-sphere model (IRHSM) is characterized by a coefficient of tangential restitution, $\beta$ \cite{JR85a,LS87,ZTPSH98,LHMZ98,CLH02,BPKZ07,CP08,SKG10,GG20}, in addition to $\alpha$. Needless to say, the NSF transport coefficients obtained from the IRHSM by an extension of the approximations used for the IHSM  are much more involved than in the latter case \cite{KSG14,GSK18,MS21a,MS21b}.

The importance of exactly solvable models in statistical physics cannot be overemphasized, as they play a pivotal role in our understanding of complex physical systems and the development of fundamental principles.
These models function as essential benchmarks, offering precise solutions that can serve as a reference for more realistic but less tractable systems
Moreover, they are invaluable for testing theoretical concepts, verifying numerical simulations, and confirming experimental observations. In this context, Maxwell models applied to the Boltzmann equation serve as essential tools in the development of the kinetic theory of gases.

As is well known, the  original Maxwell molecules  interact via a continuous repulsive interaction potential decaying as $r^{-4}$ with the intermolecular distance $r$ \cite{M67,TM80,GS03,S09b}. A notable consequence of this behavior is the independence of the collision rate on the relative velocity of the colliding pair. Consequently, the velocity moments of the Boltzmann collision term can be expressed as  bilinear combinations of moments of the distribution function of equal or lower degree, facilitating the derivation of insightful exact results.
The extension of this velocity-independence of the collision rate allows for the formulation of Maxwellian mathematical models that do not rely on a specific interaction potential \cite{E81}.

While Maxwell models were originally designed for molecular gases, the exploration of granular gases has led to the development of the inelastic Maxwell model (IMM), accompanied by the derivation of various exact results, particularly focusing on the presence of fat high-energy tails
\cite{BCG00, CCG00, BK00, C01b, BC02, KB02, BK02, BK02b, EB02a, EB02b, EB02c, BBP02, BP02a, BP02b, EB03, BK03, BC03, BCT03, SE03, BG06, ETB06a, ETB06b, BC07, GS07, BTE07, GT12b, SG12}.
Within the IMM framework, precise evaluations of the exact NSF transport coefficients
\cite{S03, GA05, CGV14, GKT15, KG15, KG20}
and the Burnett coefficients \cite{KGS14} have been conducted. Additionally, several rheological properties have been determined
\cite{G03a, SG07, G07, G08c, SGV09, GT10, GS11, GT12, GT15, GT16, GG19, RS23}.
The interest in the IMM extends beyond its mathematical tractability, encompassing its ability to account for experimental results involving magnetic grains \cite{KSSAOB05}.

Since the IMM represents a mathematical model for granular gases much more tractable than the conventional IHSM, thus allowing for exact results, it is then natural to construct a Maxwell version of the more realistic IRHSM. In this sense, we recently proposed the inelastic rough Maxwell model (IRMM) \cite{KS22} and derived the exact collisional production rates  of first and second degree, as well as the most relevant ones of third and fourth degree. In addition, the results were applied to the evaluation of the rotational-to-translational temperature ratio and the velocity cumulants in the HCS.

The aim of this paper is to derive the exact expressions for the NSF transport coefficients predicted by the IRMM, namely the shear and bulk viscosities, the thermal and diffusive heat conductivities, and the cooling-rate transport coefficient, that is, the same coefficients as previously obtained from the IRHSM by the application of a standard Sonine approximation \cite{KSG14,MS21a}. However, in contrast to the approximate treatment in the IRHSM, the study carried out here unveils an inherent coupling of the shear viscosity to a spin viscosity (associated with the spin-spin tensor defined below). Likewise, the heat conductivities are coupled to their counterparts associated with a torque-vorticity vector defined below. Thus, the exact results derived here can shed light on  improvements of the Sonine approximation as applied to the IRHSM.

Our study is motivated by the challenge of deriving exact NSF transport coefficients for granular gases, focusing on the IRMM. Unlike molecular gases, granular gases present unknown nonequilibrium base states, complicating the derivation process. Existing models rely on nontrivial approximations, but the IRMM avoids such complexities and uncovers previously overlooked couplings, thus enhancing our understanding of granular gas dynamics. This contributes to refining approximation methods in similar models, serving as a valuable benchmark in both theory and practical applications.

This paper is organized as follows. The IRMM is presented in Sect.~\ref{sec2}, where also the necessary collisional production rates are provided, the expressions of the associated coefficients being found in the Appendix. Next, in Sect.~\ref{sec3}, the Chapman--Enskog method is applied to the explicit evaluation of the NSF transport coefficients. The results are extensively discussed in Sect.~\ref{sec4}, including a stability analysis of the HCS and an assessment of the impact of the couplings mentioned above. Finally, the paper is closed in Sect.~\ref{sec5} with some concluding remarks.

\section{The inelastic rough Maxwell model. Exact collisional production rates}\label{sec2}
Let us consider a granular gas made of spherical particles of diameter $\sigma$, mass $m$, and moment of inertia $I$. The translational and angular velocities of a particle will be denoted by $\bc$ and $\bw$, respectively. According to the IRMM, the Boltzmann equation for the gas  reads \cite{KS22}
 \beq
\frac{\partial f}{\partial t}+\bc\cdot\nabla f=J[\bxi\vert f,f],
\label{15a}
\eeq
where $\bxi\equiv\{\bc,\bw\}$, $d\bxi\equiv d\bc d\bw$, $f(\bx,\bxi, t)$ is the one-particle velocity distribution function (VDF), and $J[\bxi\vert f,f]$ is the bilinear collision operator
 \beq
J[\bxi_1\vert f,f]=\frac{\nuM}{4\pi n}\int d\bxi_2\int d\bk\,\left(\frac{\hat{b}_{12}^{-1}}{\alpha\beta^2} -1\right)f(\bxi_1)f(\bxi_2).
\label{15M}
 \eeq
Here, $\nuM$   is an effective mean-field collision frequency, $n=\int d\bxi\, f(\bxi)$ is the number density, and $\bk$ is the intercenter unit vector at contact. The operator $\hat{b}_{12}$  expresses the postcollision velocities as functions of the precollisional velocities, the unit vector $\bk$, the coefficient of normal restitution $\alpha$, the coefficient of tangential restitution $\beta$, and the reduced moment of inertia $\kappa\equiv 4I/m\sigma^2$ \cite{G19,KSG14,MS21a}. For a brief description of the collision rules, the reader is referred to Sect.~2.1 of Ref.~\cite{KS22}.

The range of the coefficient of normal restitution is $0\leq\alpha\leq 1$, where $\alpha=1$ corresponds to a perfectly elastic collision (in the sense that the normal component of the relative velocity just changes its sign upon collision). In the case of the coefficient of tangential restitution, the range is $-1\leq\beta\leq 1$, where $\beta=-1$ defines smooth particles (the tangential component of the relative velocity is preserved after a collision) and $\beta=1$ corresponds to perfectly rough particles (the tangential component of the relative velocity just changes its sign upon collision). Only if  $\alpha=\mid\!\!\beta\!\!\mid=1$ is the kinetic energy conserved by collisions.
As for the reduced moment of inertia, $\kappa=0$ refers to a concentration
of the mass at the center of the sphere,  $\kappa=\frac{2}{5}$ means  a uniform mass distribution, and $\kappa=\frac{2}{3}$ if the mass is concentrated on the surface.

Given an arbitrary function $\Psi(\bxi)$, we define its average value as
\beq
\langle\Psi\rangle =\frac{1}{n}\int d\bxi\, \Psi(\bxi)f(\bxi).
\label{avpsi}
\eeq
In particular,
\begin{subequations}
\label{4a-4e}
\beq
\mathbf{u}=\medio{\bc},\quad \bW=\medio{\bw},
\eeq
\beq
T_t=\frac{m}{3}\medio{V^2},\quad T_r=\frac{I}{3}\medio{\omega^2},\quad T=\frac{1}{2}\left(T_t+T_r\right),
\eeq
\beq
 \Upsilon_{ij}=n\medio{\omega_iV_j},\quad \Pi_{ij}=n m\medio{V_iV_j-\frac{1}{3}V^2\delta_{ij}},
\eeq
\beq
  \Omega_{ij}=nI\medio{\omega_i\omega_j-\frac{1}{3}\omega^2\delta_{ij}},
  \quad \mathbf{Q}=\frac{nI}{2}\medio{(\mathbf{V}\cdot\bw)\bw},
\eeq
\beq
\mathbf{q}_t=\frac{nm}{2}\medio{V^2\mathbf{V}},\quad \mathbf{q}_r=\frac{nI}{2}\medio{\omega^2\mathbf{V}},\quad \mathbf{q}=\mathbf{q}_t+\mathbf{q}_r.
\eeq
\end{subequations}
Here, $\mathbf{u}$ is the flow velocity, $\mathbf{V}\equiv\bc-\buu$ is the translational peculiar velocity, $T_t$ is the translational  granular temperature, $\Pi_{ij}$ is the (traceless) stress tensor, and $\mathbf{q}_t$ is the translational heat flux vector. Note that the pressure tensor is
\beq
\label{Pij}
P_{ij}=nm\medio{V_iV_j}=\Pi_{ij}+nT_t\delta_{ij}.
\eeq
The quantities $T_t$, $\Pi_{ij}$, $P_{ij}$, and $\mathbf{q}_t$ are defined in the same way as in the case of smooth particles. However, the rotational degrees of freedom make it necessary to define additional quantities like the mean spin vector $\bW$, the rotational granular temperature $T_r$, the (traceless) spin-spin tensor $\Omega_{ij}$, and the rotational heat flux vector $\mathbf{q}_r$. The mean granular temperature and the total heat flux are $T$ and $\mathbf{q}$, respectively. Finally, the asymmetric quantity $\Upsilon_{ij}$ represents the couple stress tensor \cite{B97,MHN02}, while $\mathbf{Q}$ can be seen as a torque-vorticity vector.

The collisional production rate of an arbitrary function $\Psi(\bxi)$ is
\bal
\label{16J}
\mathcal{J}[\Psi]\equiv&\frac{1}{n}\int d\bxi\,\Psi(\bxi)J[\bxi\vert f,f]\nn
=& \frac{\nuM}{8\pi n^2}\int d\bxi_1\int d\bxi_2\int d\bk\,f(\bxi_1)f(\bxi_2)
\left(\hat{b}_{12}-1\right)\left[\Psi(\bxi_1)+\Psi(\bxi_2)\right].
\eal
In Ref.~\cite{KS22}, it was proved that the collisional production rates associated with the quantities defined in Eqs.~\eqref{4a-4e} are  $\mathcal{J}\left[\bc\right]=0$ and
\begin{subequations}
\label{6A-6H}
\beq
\label{1a}
 -\nuM^{-1}{\mathcal{J}\left[\bw\right]}= \varphi_{01\mid  01}  \bW,\quad
-\nuM^{-1}{\mathcal{J}\left[n\omega_i V_j\right]}= \psi_{11\mid  11} \Upsilon_{ij},
\eeq
\beq
\label{6b}
-\nuM^{-1}{\mathcal{J}\left[\frac{m}{3}V^2\right]}=\chi_{20\mid  20}T_t  +\frac{4}{\ka}\chi_{20\mid  02}T_r,
\eeq
\beq
\label{6c}
-\nuM^{-1}{\mathcal{J}\left[\frac{I}{3}\omega^2\right]}=
  \frac{\ka}{4}\chi_{02\mid  20}T_t+\chi_{02\mid  02}T_r,
\eeq
\beq
\label{6d}
-\nuM^{-1}{\mathcal{J}\left[nm\left(V_iV_j-\frac{1}{3}V^2\delta_{ij}\right)\right]}=
\psi_{20\mid  20}\Pi_{ij}  +\frac{4}{\ka}\psi_{20\mid  02}\Omega_{ij},
\eeq
\beq
-\nuM^{-1}{\mathcal{J}\left[nI\left(\omega_i\omega_j-\frac{1}{3}\omega^2\delta_{ij}\right)\right]}=
 \frac{\ka }{4}\psi_{02\mid  20}\Pi_{ij}+\psi_{02\mid  02}\Omega_{ij}   ,
\eeq
\beq
-\nuM^{-1}{\mathcal{J}\left[n\frac{m}{2}V^2\mathbf{V}\right]}=\varphi_{30\mid  30}\mathbf{q}_t+\frac{4}{\ka}\varphi_{30\mid  12}\left(2\mathbf{q}_r-\mathbf{Q}\right),
\eeq
\beq
-\nuM^{-1}{\mathcal{J}\left[n\frac{I}{2}\omega^2\mathbf{V}\right]}=\varphi_{12\mid  12}^{(1)}\mathbf{q}_r+\varphi_{12\mid  12}^{(2)}\mathbf{Q}+\frac{\ka}{4}\varphi_{12\mid  30}\mathbf{q}_t,
\eeq
\beq
\label{6h}
-\nuM^{-1}{\mathcal{J}\left[n\frac{I}{2}(\mathbf{V}\cdot\bw)\bw\right]}=\overline{\varphi}_{12\mid  12}^{(1)}\mathbf{Q}+
\overline{\varphi}_{12\mid  12}^{(2)}\mathbf{q}_r.
\eeq
\end{subequations}
It should be noted that in Eqs.~\eqref{6b}--\eqref{6h} it has been assumed that $\bW=\Upsilon_{ij}=0$ (see below).
The reasoning behind the notation of the coefficients in Eqs.~\eqref{6A-6H}, along with their explicit expressions, is detailed in the Appendix.

The cooling rate $\zeta$ is defined as
\beq
\int d\bxi \left(mV^2+I\omega^2\right)J[\bxi\vert f,f]=-6\zeta nT.
\eeq
Thus,
\beq
\label{8}
\zeta=\frac{\nuM}{2T}\left[\left(\chi_{20\mid  20}+\frac{\ka}{4}\chi_{02\mid  20}\right)T_t+\left(\frac{4}{\ka}\chi_{20\mid  02}+\chi_{02\mid  02}\right)T_r\right].
\eeq

\section{Chapman--Enskog expansion}
\label{sec3}

\subsection{General scheme}

The hydrodynamic balance equations of a granular gas are \cite{G19,KSG14,MS21a}
\begin{subequations}
\label{hydro}
\beq
\label{18}
{\cal D}_tn+n\nabla\cdot \buu=0,
\eeq
\beq
\label{19}
m n{\cal D}_t \buu
+\nabla\cdot \mathsf{P}
=0,
\eeq
\beq
\label{20}
{\cal D}_t T +{1}{3n}\left(
\nabla\cdot \mathbf{q}+\mathsf{P}:\nabla \buu\right)+ T  \zeta=0.
\eeq
\end{subequations}
In the above balance equations,  $\mathcal{D}_t=\partial_t+\buu\cdot\nabla$ denotes the material time derivative. Equations \eqref{hydro} are formally exact but they do not make a closed set unless constitutive equations expressing the pressure tensor $\mathsf{P}$, the heat flux $\mathbf{q}$, and the cooling rate $\zeta$ in terms of gradients of the hydrodynamic fields ($n$, $\buu$, and $T$) are proposed. The form of the NSF constitutive equations (to first order in gradients) is
\begin{subequations}
\label{NSF}
\beq
\label{PijNSF}
P_{ij}=n T_t^{(0)}\delta_{ij}-\eta \Delta_{ijk\ell}\nabla_k u_\ell-\eta_b\nabla\cdot\buu \delta_{ij},
\eeq
\beq
\label{qNSF}
\mathbf{q}=-\lambda\nabla T-\mu\nabla n,
\eeq
\beq
\label{xiNSF}
\zeta=\zeta^{(0)}-\xi\nabla\cdot\buu,
\eeq
\end{subequations}
where the superscript $(0)$ denotes quantities in the absence of gradients, $\eta$ is the shear viscosity, $\eta_b$ is the bulk viscosity, $\lambda$ is the thermal heat conductivity, $\mu$ is a Dufour-like coefficient that will be referred to as the diffusive heat conductivity \cite{BR04,NBSG07,G19}, and $\xi$ is a  cooling-rate transport coefficient. In Eq.~\eqref{PijNSF},
\beq
\Delta_{ijk\ell}\equiv\delta_{ik}\delta_{j\ell}+\delta_{i\ell}\delta_{jk}-\frac{2}{3}\delta_{ij}\delta_{k\ell}.
\eeq

Although the spin-spin tensor $\boldsymbol{\Omega}$, the partial heat fluxes $\mathbf{q}_t$ and $\mathbf{q}_r$, and the torque-vorticity vector
$\mathbf{Q}$ do not enter into the hydrodynamic balance equations, Eqs.~\eqref{hydro}, they are intrinsically coupled to $\mathbf{P}$ and $\mathbf{q}$, as will be seen below.

Moreover, multiplying both sides of Eq.~\eqref{15a} by $\omega_i$ and integrating over $\bxi$, one gets the following balance equation:
\beq
n{\cal D}_t\Omega_i+\nabla_j\Upsilon_{ij}=-n\nu \varphi_{01\mid  01}\Omega_i,
\eeq
where use has been made of Eq.~\eqref{1a}. This shows that, in a homogeneous state, $\partial_t\bW=-\nu \varphi_{01\mid  01}\bW$, so that $\lim_{t\to\infty}\bW(t)=0$. As we will see later, $\bW\to 0$  even in the presence of hydrodynamic gradients to NSF order.

Our main goal is to derive the  exact expressions for the NSF transport coefficients within the IRMM. To that end, we assume that the VDF depends on space and time only through the {slow} hydrodynamic fields ($n$, $\buu$, and $T$), and apply  the Chapman--Enskog  expansion method \cite{G19,KSG14,MS21a}.

In the method, one introduces a bookkeeping parameter $\epsilon$, which is used as a small parameter, so that
\begin{subequations}
\beq
 \nabla \rightarrow \epsilon \nabla,\quad    f = f^{(0)} + \epsilon f^{(1)} + \epsilon^2 f^{(2)}+\cdots, \label{eq:29a}
\eeq
\beq
    \Dert =\Dert^{(0)}+\epsilon\Dert^{(1)}+\epsilon^2\Dert^{(2)}+\cdots. \label{eq:29b}
\eeq
\end{subequations}
Thus, the Boltzmann equation, Eq.\ \eqref{15a}, decouples into a hierarchy of equations of orders $k=0,1,2,\ldots$. To make a qualitative contact with the case of hard spheres, we will assume $\nuM\propto n\sqrt{T_t}$ and, therefore,  the collision frequency  must also be expanded as
\beq
\nuM=\nuM^{(0)}+\epsilon\nuM^{(1)}+\epsilon^2 \nuM^{(2)}+\cdots,
\eeq
where
\beq
\label{nu1}
\nuM^{(0)}\propto n\sqrt{T_t^{(0)}},\quad\nuM^{(1)}=\nuM^{(0)}\frac{T_t^{(1)}}{2T_t^{(0)}}.
\eeq

\subsection{Zeroth-order distribution}
The zeroth-order Boltzmann equation is
\beq
    \Dert^{(0)} f^{(0)}(\bxi_1) =\frac{\nuM^{(0)}}{4\pi n}\int d\bxi_2\int d\bk\,\left(\frac{\hat{b}_{12}^{-1}}{\alpha\beta^2} -1\right)f^{(0)}(\bxi_1)
       f^{(0)}(\bxi_2). \label{eq:35a}
\eeq
This shows that the zeroth-order VDF $f^{(0)}$ is the local version of the HCS VDF. Thus, according to the results derived in Ref.~\cite{KS22},
\beq
\label{Tt0}
T_t^{(0)}=\tau_t T,\quad T_r^{(0)}=\tau_r T,
\eeq
where
\begin{subequations}
\label{22ab0}
\beq
\tau_t=2-\tau_r=\frac{2}{1+\theta},\quad
\theta=h+\sqrt{1+h^2},
\eeq
\bal
h\equiv&\frac{\ka}{8}\frac{\chi_{02\mid  02}-\chi_{20\mid  20}}{\chi_{20\mid  02}}\nn
=&\frac{1+\ka}{2\ka(1+\be)}\left[\frac{1+\ka}{2}\frac{1-\al^2}{1+\be}-(1-\ka)(1-\be)\right].
\eal
\end{subequations}
Moreover, the zeroth-order cooling rate is
\bal
\label{zeta0}
\zeta^{(0)}=&\frac{\nuM^{(0)}}{2}\left[\left(\chi_{20\mid  20}+\frac{\ka}{4}\chi_{02\mid  20}\right)\tau_t
+\left(\frac{4}{\ka}\chi_{20\mid  02}+\chi_{02\mid  02}\right)\tau_r\right]\nn
=&\frac{1}{6}\frac{\nuM^{(0)}}{1+\theta}\left[1-\al^2+2\frac{1-\be^2}{1+\ka}\theta\left(\frac{\ka}{\theta}+1\right)\right].
\eal
Note also that, in the HCS,
\beq
\label{bW0}
\bW^{(0)}=\Upsilon_{ij}^{(0)}=\Pi_{ij}^{(0)}=\Omega_{ij}^{(0)}=\mathbf{q}_t^{(0)}=\mathbf{q}_r^{(0)}=\mathbf{Q}^{(0)}=0.
\eeq

\subsection{First-order distribution}
The integral equation for the first-order distribution function $f^{(1)}$ is
\bal
 \label{24b}
 \left({\cal D}_t^{(0)} +\mathcal{L}\right)f^{(1)}=&-\left({\cal D}_t^{(1)}+\mathbf{V}\cdot\nabla -\frac{T_t^{(1)}}{2\tau_tT}{\cal D}_t^{(0)}\right)f^{(0)}\nn
 =&\mathbf{A}\cdot\nabla \ln T+\mathbf{B}\cdot\nabla \ln n{+C_{ij}\nabla_j u_i}
+E\nabla\cdot\buu\nn
&+\left[\zeta^{(1)}-\frac{\nuM^{(1)}\zeta^{(0)}}{\nuM^{(0)}}\right]T \partial_T f^{(0)},
 \eal
where
\bal
\mathcal{L}f^{(1)}(\bxi_1)=&-\frac{\nuM^{(0)}}{4\pi n}\int d\bxi_2\int d\bk\,\left(\frac{\hat{b}_{12}^{-1}}{\alpha\beta^2} -1\right)
\nn&\times
\left[f^{(0)}(\bxi_1)f^{(1)}(\bxi_2)
+f^{(0)}(\bxi_2)f^{(1)}(\bxi_1)\right]
\label{15L}
 \eal
is the linearized collision operator and
\begin{subequations}
\label{eqA-eqT}
\beq
\mathbf{A}=-T\left(\mathbf{V}\partial_T+\frac{\tau_t}{m}\partial_{\mathbf{V}}\right)f^{(0)},
\label{eqA}
\eeq
\beq
\mathbf{B}=-\left(\mathbf{V}+\frac{\tau_t T}{m}\partial_{\mathbf{V}}\right)f^{(0)},
\label{eqB}
\eeq
\beq
{C_{ij}=\left(\partial_{V_i}V_j-\frac{1}{3}\delta_{ij}\partial_{\mathbf{V}}\cdot\mathbf{V}\right)f^{(0)}},
\label{eqC}
\eeq
\beq
E=\frac{1}{3}\left(\partial_\mathbf{V}\cdot\mathbf{V}+\tau_tT\partial_T\right)f^{(0)},
\label{eqE}
\eeq
\beq
\label{eqT}
T\partial_T f^{(0)}=-\frac{1}{2}\left(\partial_{\bV}\cdot\bV+\partial_{\bw}\cdot\bw\right)f^{(0)}.
\eeq
\end{subequations}
The solution to Eq.\ \eqref{24b} has the form
\beq
f^{(1)}=
\bbA\cdot\nabla \ln T+\bbB\cdot\nabla \ln n+\bbC_{ij}{\nabla_j u_i}
+\bbE\nabla\cdot\buu,
\label{f1}
\eeq
where the vectors $\bbA$ and $\bbB$, the traceless tensor $\bbC_{ij}$, and the scalar $\bbE$ are the solutions of a set of linear integral equations \cite{KSG14,MS21a}. Whereas in the case of the IRHSM one needs to solve those integral equations by a Sonine approximation, the main advantage of the IRMM is that it is not necessary to get any explicit expression for $f^{(1)}$ (or, equivalently, $\bbA$, $\bbB$, $\bbC_{ij}$ and $\bbE$) since, as shown below, one can exactly obtain the transport coefficients by just taking velocity moments in both sides of Eq.~\eqref{24b}.

From Eq.~\eqref{15L}, it follows that
\beq
\label{13}
\int d\bxi\,\Psi(\bxi)\mathcal{L}f^{(1)}(\bxi)=-n\nuM^{(0)}\nuM^{-1}\mathcal{J}[\Psi]
\eeq
for the functions $\Psi(\bxi)$ in Eqs.~\eqref{1a} and \eqref{6d}--\eqref{6h}.
On the other hand, from Eqs.~\eqref{6b} and \eqref{6c}, we have
\beq
\label{1b}
\int d\bxi
\begin{Bmatrix}
  \frac{m}{3}V^2\\
  \frac{I}{3}\omega^2
\end{Bmatrix}
\mathcal{L}f^{(1)}(\bxi)=n\nuM^{(0)}
\begin{Bmatrix}
\chi_{20\mid  20}  -\frac{4}{\ka}\chi_{20\mid  02}\\
  \frac{\ka}{4}\chi_{02\mid  20}-\chi_{02\mid  02}
\end{Bmatrix}{T_t^{(1)}},
\eeq
where  we have taken into account that $T_t^{(1)}+T_r^{(1)}=0$ since $T$ is a hydrodynamic variable.
{}From Eq.~\eqref{8}, we obtain the following expression for the  first-order cooling rate,
\beq
\label{2}
\zeta^{(1)}=\nuM^{(0)}\left[\frac{\zeta^{(0)}}{\tau_t\nuM^{(0)}}+\chi_{20\mid  20}  -\frac{4\chi_{20\mid  02}}{\ka}-\chi_{02\mid  02} +\frac{\ka\chi_{02\mid  20}}{4}\right]\frac{T_t^{(1)}}{2T},
\eeq
where we have taken into account that $\nuM^{(1)}/\nuM^{(0)}={T_t^{(1)}}/{2\tau_t T}$ [see Eq.~\eqref{nu1}].

As we will show below, Eqs.~\eqref{13} and \eqref{1b}, together with the collision integrals in Eqs.~\eqref{6A-6H}, allow us to derive exactly the NSF transport coefficients within the IRMM.

\subsubsection{Mean spin vector and couple stress tensor}

Here we prove that $\bW^{(1)}=\Upsilon_{ij}^{(1)}=0$.
Suppose that  $\bW^{(1)}$ and $\Upsilon_{ij}^{(1)}$ existed; then, by symmetry, one should have
\begin{subequations}
\beq
n\bW^{(1)}=\varsigma_1\nabla T+\varsigma_2\nabla n,
\eeq
\beq
\label{Upsilon1}
\Upsilon_{ij}^{(1)}=\varsigma_{ijk\ell}\nabla_ku_\ell,
\eeq
\end{subequations}
where $\varsigma_1$, $\varsigma_2$, and $\varsigma_{ijk\ell}$ would be the associated transport coefficients. By dimensional analysis, $\varsigma_1\propto n/\sqrt{mT}$, $\varsigma_2\propto\sqrt{T/m}$, $\varsigma_{ijk\ell}\propto n\sqrt{T/m}$. Therefore,
\begin{subequations}
\label{19ab}
\beq
\label{19a}
\mathcal{D}_t^{(0)}n\bW^{(1)}=\zeta^{(0)}\left[-\varsigma_1\nabla T+\left(\frac{\varsigma_1 T}{n}-\varsigma_2\right)\nabla n\right],
\eeq
\beq
\mathcal{D}_t^{(0)}\Upsilon_{ij}^{(1)}=-\frac{\zeta^{(0)}}{2}\varsigma_{ijk\ell}\nabla_ku_\ell.
\eeq
\end{subequations}
In Eq.~\eqref{19a}, use has been made of the property
\beq
\mathcal{D}_t^{(0)}\nabla T=-\nabla\left[\zeta^{(0)}T\right]=-\zeta^{(0)}\left(\frac{T}{n}\nabla n+\frac{3}{2}\nabla T\right).
\eeq

The key point is that, as can be easily checked from Eqs.~\eqref{eqA-eqT},
\bal
\int d\bxi
\begin{Bmatrix}
\bw\\
\omega_iV_j
\end{Bmatrix}
A_k(\bxi)=&\int d\bxi
\begin{Bmatrix}
\bw\\
\omega_iV_j
\end{Bmatrix}
B_k(\bxi)
=\int d\bxi
\begin{Bmatrix}
\bw\\
\omega_iV_j
\end{Bmatrix}
C_{k\ell}(\bxi)\nn
=&\int d\bxi
\begin{Bmatrix}
\bw\\
\omega_iV_j
\end{Bmatrix}
E(\bxi)
=\int d\bxi
\begin{Bmatrix}
\bw\\
\omega_iV_j
\end{Bmatrix}
\partial_T f^{(0)}(\bxi)=0.
\eal
As a consequence, multiplying both sides of Eq.~\eqref{24b} by $\bw$ or $\omega_iV_j$, integrating over $\bxi$, and using Eqs.\ \eqref{1a}, \eqref{13}, and \eqref{19ab}, one gets
\begin{subequations}
\beq
\varsigma_1\zeta^{(0)}=\varsigma_1\nuM^{(0)} \varphi_{01\mid  01},
\eeq
\beq
\left(\varsigma_2-\frac{\varsigma_1 T}{n}\right)\zeta^{(0)}=\varsigma_2\nuM^{(0)} \varphi_{01\mid  01},
\eeq
\beq
\varsigma_{ijk\ell}\frac{\zeta^{(0)}}{2}=\varsigma_{ijk\ell}\nuM^{(0)} \psi_{11\mid  11}.
\eeq
\end{subequations}
The solution of those homogeneous linear equations is the trivial one, i.e.,
\beq
\label{varsigma}
\varsigma_1=\varsigma_2=\varsigma_{ijk\ell}=0.
\eeq
Therefore, $\bW^{(1)}=\Upsilon_{ij}^{(1)}=0$, thus justifying the assumption made in Eqs.~\eqref{6b}--\eqref{6h}.

The observation that both the mean spin vector and the couple stress tensor vanish to  NSF order in the IRMM contrasts with expectations based on the micropolar fluids framework \cite{E66, B97, L99b, MHN02}. This contrast is likely attributed to the low-density regime where the Boltzmann equation is applicable and the absence of boundary effects in the NSF description \cite{MHN02}. However, in the case of dense gases, a dependency does emerge, with the stress tensor correlating with the anti-symmetric part of flow velocity and mean spin gradients, and the heat flux linking with the curl of the mean spin vector \cite{MSD66,DT75,GK91,K24}.

\subsubsection{Stress and spin-spin tensors}
Now we turn to the first-order translational temperature $T_t^{(1)}$, stress tensor $\Pi_{ij}^{(1)}$, and spin-spin tensor $\Omega_{ij}^{(1)}$. First, we need to make use of the integrals
\begin{subequations}
\beq
\int d\bxi\, \left\{
\begin{array}{c}
m V_iV_j\\
I \omega_i\omega_j
\end{array}
\right\} \mathbf{A}(\bxi)=\int d\bxi\, \left\{
\begin{array}{c}
mV_iV_j\\
I\omega_i\omega_j
\end{array}
\right\} \mathbf{B}(\bxi)=0,
\eeq
\beq
\int d\bxi\, \left\{
\begin{array}{c}
mV_iV_j\\
I\omega_i\omega_j
\end{array}
\right\} {E}(\bxi)=\frac{n\tau_t\tau_rT}{3}\left\{
\begin{array}{c}
-1\\
1
\end{array}
\right\}\delta_{ij},
\eeq
\beq
\int d\bxi\, \left\{
\begin{array}{c}
mV_iV_j\\
I\omega_i\omega_j
\end{array}
\right\} T\partial_T f^{(0)}(\bxi)={nT}\left(
\begin{array}{c}
\tau_t\\
\tau_r
\end{array}
\right)\delta_{ij},
\eeq
\beq
\int d\bxi\, \left\{
\begin{array}{c}
mV_iV_j\\
I\omega_i\omega_j
\end{array}
\right\} C_{k\ell}(\bxi)=-{n\tau_tT}\left\{
\begin{array}{c}
\Delta_{ijk\ell}\\
0
\end{array}
\right\}.
\eeq
\end{subequations}
Multiplying both sides of Eq.~\eqref{24b} by $mV^2/3$ and integrating over $\bxi$, one gets
\bal
\label{3a}
\mathcal{D}_t^{(0)}
 T_t^{(1)}
+\nuM^{(0)}
\left(\chi_{20\mid  20}  -\frac{4}{\ka}\chi_{20\mid  02}\right){T_t^{(1)}}
=&
-\frac{\tau_t \tau_r}{3} T\nabla\cdot\buu\nn
&+\left[\zeta^{(1)}-\frac{\nuM^{(1)}\zeta^{(0)}}{\nuM^{(0)}}\right]
\tau_t  T.
\eal
The same result is obtained by multiplication of both sides of Eq.~\eqref{24b} by $I\omega^2/3$.

Since, on physical grounds, $T_t^{(1)}$ must be proportional to $ T(\nabla\cdot\buu)/\nuM^{(0)}$, it turns out that $\mathcal{D}_t^{(0)}T_t^{(1)}=\frac{1}{2}[T_t^{(1)}/T]\mathcal{D}_t^{(0)}T=-\frac{1}{2}\zeta^{(0)}T_t^{(1)}$.
Thus, $T_t^{(1)}$ is given by
\beq
\label{Tt1}
{T_t^{(1)}}=-\frac{\eta_b}{n} \nabla\cdot\buu,
\eeq
where
\beq
\label{eta_b}
\eta_b=\frac{2nT\tau_t\tau_r}{3\nuM^{(0)}}\left[\left(\chi_{20\mid  20}  -\frac{4\chi_{20\mid  02}}{\ka}\right)\tau_r
+\left(\chi_{02\mid  02}  -\frac{\ka\chi_{02\mid  20}}{4}\right)\tau_t-\frac{\zeta^{(0)}}{\nuM^{(0)}}\right]^{-1}
\eeq
is the bulk viscosity introduced in Eq.~\eqref{PijNSF}.
Combination of Eqs.~\eqref{2}, \eqref{Tt1}, and \eqref{eta_b} yields the cooling-rate transport coefficient introduced in Eq.~\eqref{xiNSF}, namely
\beq
\label{xi}
\xi=\left[\frac{\zeta^{(0)}}{\tau_t\nuM^{(0)}}+\chi_{20\mid  20}  -\frac{4\chi_{20\mid  02}}{\ka}-\chi_{02\mid  02} +\frac{\ka\chi_{02\mid  20}}{4}\right]\frac{\eta_b}{2nT/\nuM^{(0)}}.
\eeq

Next, from Eq.~\eqref{24b} one  obtains
\beq
\label{22ab}
\mathcal{D}_t^{(0)}
\begin{Bmatrix}
 \Pi_{ij}^{(1)}\\
 \Omega_{ij}^{(1)}
\end{Bmatrix}
+\nuM^{(0)}
\begin{Bmatrix}
\psi_{20\mid  20}\Pi_{ij}^{(1)} +\frac{4}{\ka}\psi_{20\mid  02}\Omega_{ij}^{(1)}\\
\psi_{02\mid  02}\Omega_{ij}^{(1)}  +\frac{\ka }{4}\psi_{02\mid  20}\Pi_{ij}^{(1)}
\end{Bmatrix}
=\nabla_k u_\ell\begin{Bmatrix}
-n\tau_t T\Delta_{ijk\ell}\\
0
\end{Bmatrix}.
\eeq
Both $\Pi_{ij}^{(1)}$ and $\Omega_{ij}^{(1)}$ must be proportional to $n T\Delta_{ijk\ell}\nabla_k u_\ell$, so that $\mathcal{D}_t^{(0)}\Pi_{ij}^{(1)}=-\frac{1}{2}\zeta^{(0)}\Pi_{ij}^{(1)}$ and $\mathcal{D}_t^{(0)}\Omega_{ij}^{(1)}=-\frac{1}{2}\zeta^{(0)}\Omega_{ij}^{(1)}$. We then write
\beq
\label{Piij1}
\Pi_{ij}^{(1)}=-\eta \Delta_{ijk\ell}\nabla_k u_\ell,\quad \Omega_{ij}^{(1)}=-\eta_\Omega \Delta_{ijk\ell}\nabla_k u_\ell,
\eeq
where $\eta$ is the shear viscosity introduced in Eq.~\eqref{PijNSF} and $\eta_\Omega$ is a new transport coefficient here called {spin viscosity}.
Inserting this into Eq.~\eqref{22ab}, we obtain the set of linear equations
\beq
\label{23ab}
\begin{bmatrix}
 \nuM^{(0)}\psi_{20\mid  20}-\frac{1}{2}\zeta^{(0)}&\nuM^{(0)}\frac{4\psi_{20\mid  02}}{\ka}\\
 \nuM^{(0)}\frac{\ka \psi_{02\mid  20}}{4}&\nuM^{(0)}\psi_{02\mid  02}-\frac{1}{2}\zeta^{(0)}
\end{bmatrix}
\cdot \begin{bmatrix}
\eta\\
\eta_\Omega
\end{bmatrix}
=\begin{bmatrix}
n\tau_t T\\
0
\end{bmatrix},
\eeq
whose solution is
\begin{subequations}
\label{etaOmega&eta}
\beq
\label{etaOmega}
\eta_\Omega=-\frac{\ka}{4}\frac{\psi_{02\mid  20}}{\psi_{02\mid  02}-\frac{1}{2}\zeta^{(0)}/\nuM^{(0)}}\eta,
\eeq
\beq
\label{eta}
\eta=\frac{n\tau_t T}{\nuM^{(0)}}\left[\psi_{20\mid  20}-\frac{1}{2}\frac{\zeta^{(0)}}{\nuM^{(0)}}-\frac{\psi_{20\mid  02}\psi_{02\mid  20}}{\psi_{02\mid  02}-\frac{1}{2}\zeta^{(0)}/\nuM^{(0)}}\right]^{-1}.
\eeq
\end{subequations}

\subsubsection{Heat flux and torque-vorticity vectors}
Let us now consider the first-order translational ($\mathbf{q}_t$) and rotational ($\mathbf{q}_r$) contributions to the heat flux ($\mathbf{q}=\mathbf{q}_t+\mathbf{q}_r$), as well as the torque-vorticity vector $\mathbf{Q}$.

The needed integrals are
\begin{subequations}
\beq
\int d\bxi\, \left\{
\begin{array}{c}
\frac{m}{2} V^2{V_i}\\
\frac{I}{2} \omega^2{V_i}\\
\frac{I}{2} (\mathbf{V}\cdot\bw)\omega_i
\end{array}
\right\} {A_j}(\bxi)=-\frac{nT^2\tau_t}{2m}\left\{
\begin{array}{c}
5\tau_t\left(1+2a_{20}\right)\\
3\tau_r\left(1+2a_{11}\right)\\
\tau_r\left(1+2b_{00}\right)
\end{array}
\right\}\delta_{ij},
\eeq
\beq
\int d\bxi\, \left\{
\begin{array}{c}
\frac{m}{2} V^2{V_i}\\
\frac{I}{2} \omega^2{V_i}\\
\frac{I}{2} (\mathbf{V}\cdot\bw)\omega_i
\end{array}
\right\} {B_j}(\bxi)=-\frac{nT^2\tau_t}{2m}\left\{
\begin{array}{c}
5\tau_t a_{20}\\
3\tau_r a_{11}\\
\tau_r b_{00}
\end{array}
\right\}\delta_{ij},
\eeq
\bal
\int d\bxi\, \left\{
\begin{array}{c}
\frac{m}{2} V^2\mathbf{V}\\
\frac{I}{2} \omega^2\mathbf{V}\\
\frac{I}{2} (\mathbf{V}\cdot\bw)\bw
\end{array}
\right\} {E}(\bxi)
=&
\int d\bxi\, \left\{
\begin{array}{c}
\frac{m}{2} V^2\mathbf{V}\\
\frac{I}{2} \omega^2\mathbf{V}\\
\frac{I}{2} (\mathbf{V}\cdot\bw)\bw
\end{array}
\right\} T\partial_T f^{(0)}(\bxi)\nn
=&\int d\bxi\, \left\{
\begin{array}{c}
\frac{m}{2} V^2\mathbf{V}\\
\frac{I}{2} \omega^2\mathbf{V}\\
\frac{I}{2} (\mathbf{V}\cdot\bw)\bw
\end{array}
\right\} C_{k\ell}(\bxi)=0.
\eal
\end{subequations}
Here,
\beq
\label{cumu}
a_{20}=\frac{3}{5}\frac{\medio{V^4}}{\medio{V^2}^2}-1,\quad
 a_{11}=\frac{\medio{V^2\omega^2}}{\medio{V^2}\medio{\omega^2}}-1,
 \quad b_{00}=3\frac{\medio{(\bV\cdot\bw)^2}}{\medio{V^2}\medio{\omega^2}}-1,
\eeq
are cumulants associated with $f^{(0)}$. Their expressions within the IRMM can be found in Ref.~\cite{KS22}.\footnote{Note that, in  the notation of Ref.~\cite{KS22}, $a_{20}=a_{20}^{(0)}$, $a_{11}=a_{11}^{(0)}$, and $b_{00}=\frac{5}{2}a_{00}^{(1)}+a_{11}^{(0)}$.} The cumulants are finite only if the coefficient of tangential restitution $\beta$ is larger than a certain $\alpha$-dependent threshold value $\beta_0(\alpha)$ \cite{KS22}.

Now, we multiply both sides of Eq.~\eqref{24b} by $\{\frac{m}{2} V^2\mathbf{V},\frac{I}{2} \omega^2\mathbf{V},\frac{I}{2} (\mathbf{V}\cdot\bw)\bw\}$, and integrate over velocity. The result is
\bal
\label{25}
\mathcal{D}_t^{(0)}\begin{Bmatrix}
\mathbf{q}_t^{(1)}\\
\mathbf{q}_r^{(1)}\\
\mathbf{Q}^{(1)}
\end{Bmatrix}
+&
\nuM^{(0)}
\begin{Bmatrix}
\varphi_{30\mid  30}\mathbf{q}_t^{(1)}+\frac{4}{\ka}\varphi_{30\mid  12}
[2\mathbf{q}_r^{(1)}-\mathbf{Q}^{(1)}
]\\
\frac{\ka}{4}\varphi_{12\mid  30}\mathbf{q}_t^{(1)}+\varphi_{12\mid  12}^{(1)}\mathbf{q}_r^{(1)}+\varphi_{12\mid  12}^{(2)}\mathbf{Q}^{(1)}\\
\overline{\varphi}_{12\mid  12}^{(1)}\mathbf{Q}^{(1)}+\overline{\varphi}_{12\mid  12}^{(2)}\mathbf{q}_r^{(1)}
\end{Bmatrix}\nn
&=
-\frac{nT\tau_t}{2m}\left\{
\begin{array}{c}
5\tau_t\left(1+2a_{20}\right)\\
3\tau_r\left(1+2a_{11}\right)\\
\tau_r\left(1+2b_{00}\right)
\end{array}
\right\}\nabla T
-\frac{T^2\tau_t}{2m}\left\{
\begin{array}{c}
5\tau_t a_{20}\\
3\tau_r a_{11}\\
\tau_r b_{00}
\end{array}
\right\}\nabla n.
\eal
The solution to Eq.~\eqref{25} has the structure
\beq
\label{lambdaQmuQ}
\begin{Bmatrix}
\mathbf{q}_t^{(1)}\\
\mathbf{q}_r^{(1)}\\
\mathbf{Q}^{(1)}
\end{Bmatrix}=-\begin{Bmatrix}
\tau_t\lambda_t\\
\tau_r\lambda_r\\
\lambda_Q
\end{Bmatrix}
\nabla T
-\begin{Bmatrix}
\mu_t\\
\mu_r\\
\mu_Q
\end{Bmatrix}
\nabla n,
\eeq
where, by dimensional analysis, $\{\lambda_t,\lambda_r,\lambda_Q\}\propto nT/\nuM^{(0)}$ and $\{\mu_t,\mu_r,\mu_Q\}\propto T^2/\nuM^{(0)}$. As a consequence,
\begin{subequations}
\beq
\mathcal{D}_t^{(0)}\begin{Bmatrix}
\lambda_t\\
\lambda_r\\
\lambda_Q
\end{Bmatrix}
\nabla T=-\begin{Bmatrix}
\lambda_t\\
\lambda_r\\
\lambda_Q
\end{Bmatrix}\zeta^{(0)}\left(2\nabla T+\frac{T}{n}\nabla n\right),
\eeq
\beq
\mathcal{D}_t^{(0)}\begin{Bmatrix}
\mu_t\\
\mu_r\\
\mu_Q
\end{Bmatrix}
\nabla n=
-\frac{3}{2}\begin{Bmatrix}
\mu_t\\
\mu_r\\
\mu_Q
\end{Bmatrix}\zeta^{(0)}\nabla n.
\eeq
\end{subequations}
Equation \eqref{25} then yields the matrix equations
\begin{subequations}
\label{37&38}
\beq
  \label{37}
  \begin{bmatrix}
   \tau_t\lambda_t\\
   \tau_r\lambda_r\\
   \lambda_Q\\
 \end{bmatrix}=\frac{nT\tau_t}{2m}\left(\nuM^{(0)}\mathsf{\Phi}-2\zeta^{(0)}\mathsf{I}\right)^{-1}
 \cdot
  \begin{bmatrix}
  5\tau_t\left(1+2a_{20}\right)\\
3\tau_r\left(1+2a_{11}\right)\\
\tau_r\left(1+2b_{00}\right)
 \end{bmatrix},
 \eeq
\beq
  \label{38}
 \begin{bmatrix}
   \mu_t\\
   \mu_r\\
   \mu_Q\\
 \end{bmatrix}= \frac{T}{n}\left(\nuM^{(0)}\mathsf{\Phi}-\frac{3}{2}\zeta^{(0)}\mathsf{I}\right)^{-1}
  \cdot
 \left(
 \frac{n\tau_t T}{2m}
 \begin{bmatrix}
 5\tau_t a_{20}\\
3\tau_r a_{11}\\
\tau_r b_{00}
 \end{bmatrix}
  +\zeta^{(0)}
 \begin{bmatrix}
   \tau_t\lambda_t\\
   \tau_r\lambda_r\\
   \lambda_Q\\
 \end{bmatrix}
 \right),
 \eeq
\end{subequations}
where $\mathsf{I}$ is the $3\times3$ unit matrix and
\beq
\mathsf{\Phi}\equiv \begin{bmatrix}
   \varphi_{30\mid  30}&\frac{8}{\kappa}\varphi_{30\mid  12}&-\frac{4}{\kappa} \varphi_{30\mid  12}\\
 \frac{\kappa}{4} \varphi_{12\mid  30}&  \varphi_{12\mid  12}^{(1)}& \varphi_{12\mid  12}^{(2)}\\
 0& \overline{\varphi}_{12\mid  12}^{(2)}& \overline{\varphi}_{12\mid  12}^{(1)}
 \end{bmatrix}.
 \eeq
Equation~\eqref{37}  provides the {thermal heat conductivities} $\lambda_t$ and $\lambda_r$,  as well as the thermal torque-vorticity conductivity $\lambda_Q$. Then, the {diffusive (heat and torque-vorticity) conductivities} $\mu_t$, $\mu_r$,  and $\mu_Q$ are obtained from Eq.~\eqref{38}.

\section{Discussion}
\label{sec4}

\subsection{Summary of final expressions}

To summarize, in the NSF constitutive equations, Eqs.~\eqref{NSF}, the zeroth-order translational temperature and cooling rate are given by Eqs.~\eqref{Tt0} and \eqref{zeta0}, respectively, the bulk and shear viscosities are given by Eqs.~\eqref{eta_b} and \eqref{eta}, respectively, and the cooling-rate transport coefficient $\xi$ is obtained  from Eq.~\eqref{xi}. The thermal and diffusive heat conductivities are  $\lambda=\tau_t\lambda_t+\tau_r\lambda_r$  and $\mu=\mu_t+\mu_r$, respectively, where $\lambda_t$ and $\lambda_r$ are given by Eq.~\eqref{37}, whereas $\mu_t$ and $\mu_r$ are obtained from Eq.~\eqref{38}. In Eqs.~\eqref{37&38}, the zeroth-order cumulants ($a_{20}$,  $a_{11}$, and $b_{00}$) have exact explicit expressions \cite{KS22}.
In all those results, the relevant coefficients $\chi_{ij\mid k\ell}$, $\psi_{ij\mid k\ell}$, $\varphi_{ij\mid k\ell}$, $\varphi_{12\mid 12}^{(1)}$, $\varphi_{12\mid 12}^{(2)}$, $\overline{\varphi}_{12\mid 12}^{(1)}$, and $\overline{\varphi}_{12\mid 12}^{(2)}$ can be found in the Appendix.

It is noteworthy  that the first-order stress tensor $\mathsf{\Pi}^{(1)}$ turns out to be intrinsically coupled to the first-order spin-spin tensor $\mathsf{\Omega}^{(1)}$, as Eq.~\eqref{22ab} shows. Thus, the derivation of the shear viscosity $\eta$ requires the parallel derivation of the spin viscosity $\eta_\Omega$ introduced in Eq.~\eqref{Piij1}, this new transport coefficient being given by Eq.~\eqref{etaOmega}. This remark is important because the coupling between $\mathsf{\Pi}^{(1)}$ and $\mathsf{\Omega}^{(1)}$ is usually ignored in the approximate derivation of $\eta$ for the IRHSM, where a standard Sonine approximation is employed \cite{KSG14,MS21a}.

Analogously, Eq.~\eqref{25} shows that the first-order partial heat fluxes $\mathbf{q}_t^{(1)}$ and $\mathbf{q}_r^{(1)}$ are intrinsically coupled to the first-order torque-vorticity vector $\mathbf{Q}^{(1)}$. The latter vector is characterized by the transport coefficients $\lambda_Q$ and $\mu_Q$ [see Eq.~\eqref{lambdaQmuQ}], which are obtained from Eqs.~\eqref{37} and \eqref{38}, respectively. While the coupling between $\mathbf{q}_t^{(1)}$ and $\mathbf{q}_r^{(1)}$ is taken into account in the approximate derivation of $\lambda$ and $\mu$ for the IRHSM \cite{KSG14,MS21a}, the coupling to $\mathbf{Q}^{(1)}$ is not accounted for.

In order to nondimensionalize the coefficients, it is convenient to take as a reference the shear viscosity and thermal conductivity of a gas made
of elastic and smooth spheres at the same translational temperature as that of the HCS of the granular gas. More specifically, we take
\beq
\eta_0= \frac{5}{2}\frac{n\tau_tT}{\nuM^{(0)}},\quad \lambda_0=\frac{15}{4}\frac{\tau_t\eta_0}{m}.
\eeq
Thus, the dimensionless transport coefficients are
\begin{subequations}
\beq
\eta^*=\frac{\eta}{\eta_0},\quad \eta_b^*=\frac{\eta_b}{\eta_0},\quad \eta_\Omega^*=\frac{\eta_\Omega}{\eta_0},
\eeq
\beq
\lambda^*=\frac{\lambda}{\lambda_0},\quad \mu^*=\frac{\mu}{(T/n)\lambda_0},\quad \lambda_Q^*=\frac{\lambda_Q}{\lambda_0},\quad \mu_Q^*=\frac{\mu_Q}{(T/n)\lambda_0}.
\eeq
\end{subequations}
Note that the cooling-rate transport coefficient $\xi$ is dimensionless by construction.

 \begin{table}
\begin{center}
   \caption{Special limiting cases. The polynomials in the expression of $\lambda^*$ in the Pidduck model are $\mathcal{P}_3(\kappa)=75 (14 + 125 \ka + 80 \ka^2 + 225 \ka^3)$ and $\mathcal{P}_4(\kappa)=2 (689 + 4850 \ka + 8630 \ka^2 + 5850 \ka^3 + 1125 \ka^4)$.}\label{table2}
\begin{tabular}{cccc}
\toprule
Coefficient&Pure smooth (IMM)&Quasismooth limit&Pidduck model  \\
&($\be=-1$)&($\be\to -1$)&($\alpha=\beta=1$) \\
\midrule
$\displaystyle{\eta^*}$&$\displaystyle{\frac{24}{(1+\al)(11+\al)}}$&$\displaystyle{\frac{6}{(1+\al)(4-\al)}}$&
$\displaystyle{\frac{(1+\ka)^2(3+10\ka)}{(1+5\ka)(3+5\ka)}}$\\
$\displaystyle{\eta_b^*}$&$0$&$\displaystyle{\frac{8}{5(1-\al^2)}}$&$\displaystyle{\frac{(1+\ka)^2}{20\ka}}$\\
$\displaystyle{\xi}$&$0$&$0$&$\displaystyle{0}$\\
$\displaystyle{\eta_\Omega^*}$&$0$&$0$&$\displaystyle{-\frac{5\ka(1+\ka)^2}{(1+5\ka)(3+5\ka)}}$\\
$\displaystyle{\lambda^*}$&$\displaystyle{\frac{16(17-18\alpha+9\alpha^2)}{(1+\al)(9\al-1)(5+6\alpha-3\alpha^2)}}$&
Diverging&$\displaystyle{\frac{\mathcal{P}_4(\ka)}{\mathcal{P}_3(\ka)}}$\\
$\displaystyle{\mu^*}$&$\displaystyle{\frac{16(1-\al)(41-30\alpha+9\alpha^2)}{(1+\al)^2(9\al-1)(5+6\alpha-3\alpha^2)}}$
&Diverging&$0$\\
$\displaystyle{\lambda_Q^*}$&$0$&Diverging&$\displaystyle{\frac{4}{75}}$\\
$\displaystyle{\mu^*_Q}$&$0$&Diverging&$0$\\
\botrule
\end{tabular}
\end{center}
 \end{table}

\subsection{Special limits}

The (reduced) transport coefficients $\eta^*$, $\eta_b^*$, $\xi$, $\eta_\Omega^*$, $\lambda^*$, $\mu^*$, $\lambda_Q^*$, and $\mu_Q^*$ are displayed in Table \ref{table2} for some special limits: pure smooth particles ($\beta=-1$),  the quasi-smooth limit ($\beta\to -1$), and the conservative Pidduck case ($\alpha=\beta=1$). In the case of pure smooth particles,  the rotational degrees of freedom are irrelevant and the results are obtained by formally setting $\theta\to 0$. This  allows us to recover previous results \cite{S03}.
In the quasismooth limit, the zeroth-order cumulants diverge, what produces the divergence of the transport coefficients associated with the heat
flux and the torque-vorticity vector, whereas those associated with the pressure and spin-spin tensors take finite values. Finally,  in the Pidduck model \cite{P22}, the total kinetic energy is conserved by collisions, the model having been used for polyatomic gases \cite{P22,MSD66,CC70,K10a,CLD65}. Even in that case, the couplings  $\mathsf{\Pi}\leftrightarrow\mathsf{\Omega}$ and $\mathbf{q}\leftrightarrow\mathbf{Q}$ still hold, as made evident by the nonzero values of $\eta_\Omega^*$ and $\lambda_Q^*$, respectively. Interestingly, $\lambda_Q^*$ is independent of the reduced moment of inertia $\ka$ in the Pidduck model.

\begin{figure}
\centering
\includegraphics[height=0.45\columnwidth]{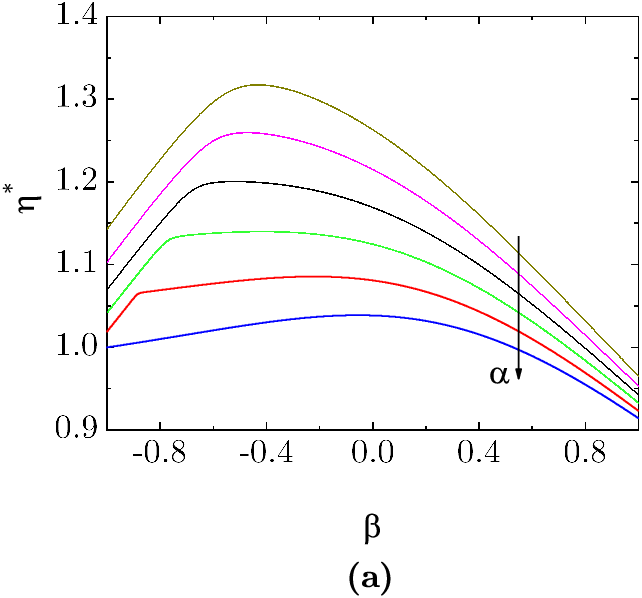}
\includegraphics[height=0.45\columnwidth]{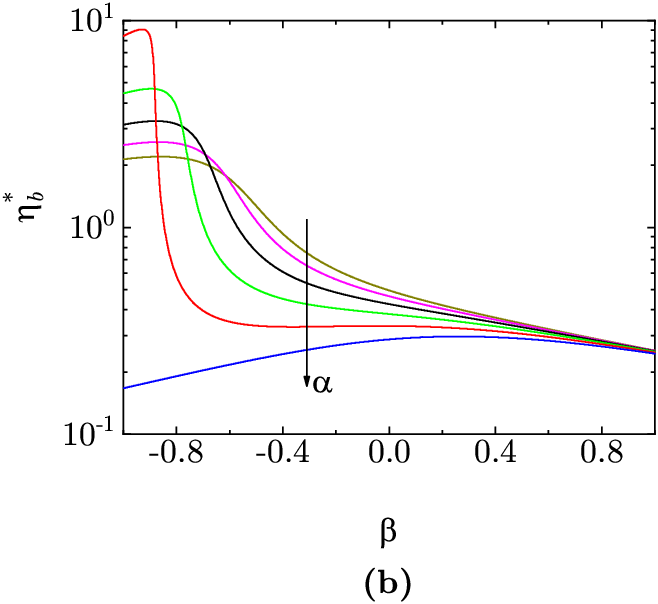}\\
\includegraphics[height=0.45\columnwidth]{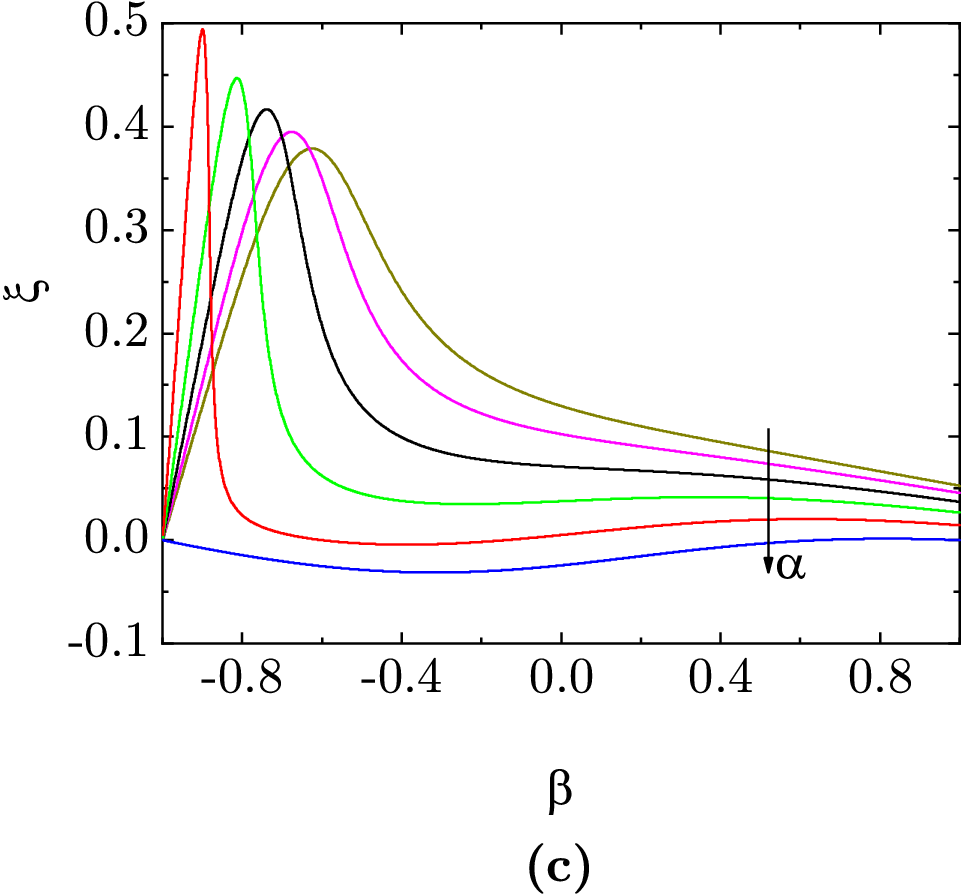}
\includegraphics[height=0.45\columnwidth]{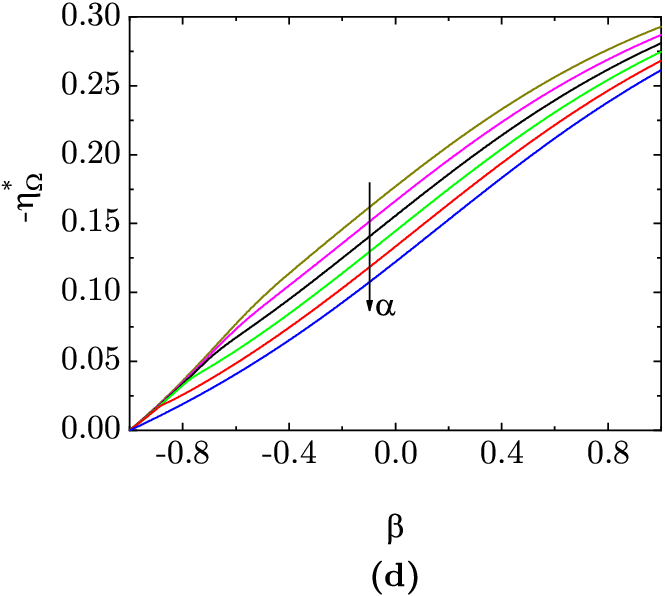}
  \caption{Plot of \textbf{a} the reduced shear viscosity $\eta^*$, \textbf{b} the reduced bulk viscosity $\eta_b^*$,  \textbf{c} the cooling-rate transport coefficient $\xi$, and \textbf{d} the reduced spin viscosity $-\eta_\Omega^*$ versus the coefficient of tangential restitution $\beta$ for $\alpha=0.5, 0.6,0.7,0.8,0.9,1$. The arrows indicate increasing values of $\alpha$. Here, the particles are assumed to have a uniform mass distribution ($\kappa=\frac{2}{5})$.}
  \label{fig1}
\end{figure}

\subsection{Coefficients $\eta^*$, $\eta_b^*$, $\xi$, and $\eta_\Omega^*$}

The NSF transport coefficients related to the pressure tensor,  the cooling rate, and the spin-spin tensor are plotted in Fig.~\ref{fig1} as functions of the coefficient of tangential restitution for several values of the coefficient of normal restitution. Here, and in what follows, we assume a uniform mass distribution in each particle, i.e., $\ka=\frac{2}{5}$.

Figures \ref{fig1}a and \ref{fig1}b show a dependence of $\eta^*$ and $\eta_b^*$ on $\alpha$ and $\beta$ qualitatively similar to that observed in the case of the IRHSM within the standard Sonine approximation \cite{KSG14}. For a fixed $\alpha$,  $\eta^*$  exhibits a local maximum at a certain value of $\beta$. As $\alpha$ decreases, the maximum grows and moves toward smaller values of $\beta$. In the case of $\eta_b^*$, however, the magnitude of the maximum decreases with decreasing $\alpha$, with the exception of $\alpha=1$. Moreover, as roughness increases, $\eta_b^*$ becomes less and less dependent on $\alpha$. In what concerns the cooling-rate transport coefficient $\xi$, Fig.~\ref{fig1}c presents a nonmonotonic dependence on both $\alpha$ and $\beta$ similar to that observed in the IRHSM with the standard Sonine approximation \cite{KSG14}.

The new transport coefficient $\eta_\Omega^*$, which is discarded in the standard Sonine approximation for the IRHSM, is displayed in Fig.~\ref{fig1}d. The most remarkable feature is that, in contrast to $\eta^*$, $\eta_\Omega^*$ is negative. To better grasp the consequences of that, imagine the simple shear flow geometry $\nabla_i u_j=\dot{\gamma}\delta_{iy}\delta_{jx}$, where the shear rate $\dot{\gamma}$ is assumed to be positive. Then, according to Eq.~\eqref{PijNSF}, $\medio{V_x V_y}<0$, that is, particles moving with $V_y>0$ tend to have $V_x<0$, and vice versa (second and fourth quadrants on the $xy$ plane). On the other hand, Eq.~\eqref{Piij1} implies that $\medio{\omega_x\omega_y}>0$, so that the projection on the $xy$ plane of the angular velocity of the particles tends to lie on the first and third quadrants. This is consistent with the orthogonality condition $\medio{\bV\cdot\bw}=0$ implied by $\Upsilon_{ij}^{(1)}=0$ [cf.\ Eqs.~\eqref{Upsilon1} and \eqref{varsigma}].

Figure~\ref{fig1}d shows that the magnitude of $\eta_\Omega^*$ grows quasilinearly with $\beta$. Moreover, at fixed $\beta$, $\mid\!\eta_\Omega^*\!\mid$ increases with decreasing $\alpha$. All of this represents a smoother and more regular dependence on $\alpha$ and $\beta$ than in the cases of the coefficients $\eta$, $\eta_b$, and $\xi$.

\begin{figure}
\centering
\includegraphics[height=0.45\columnwidth]{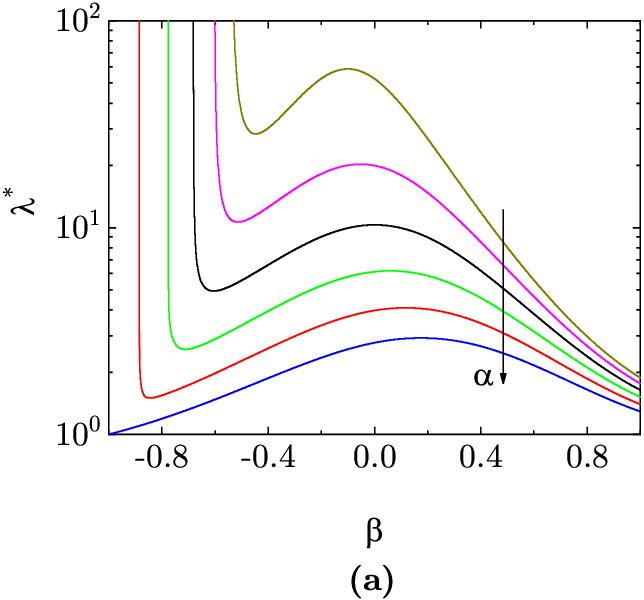}
\includegraphics[height=0.45\columnwidth]{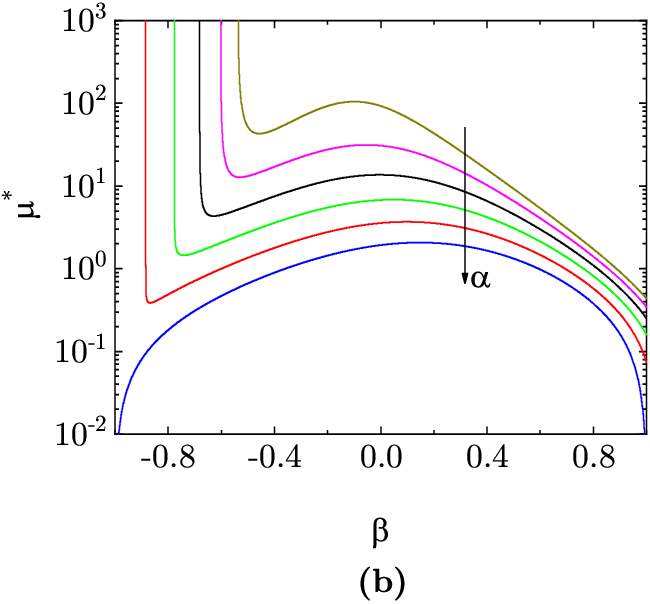}\\
\includegraphics[height=0.45\columnwidth]{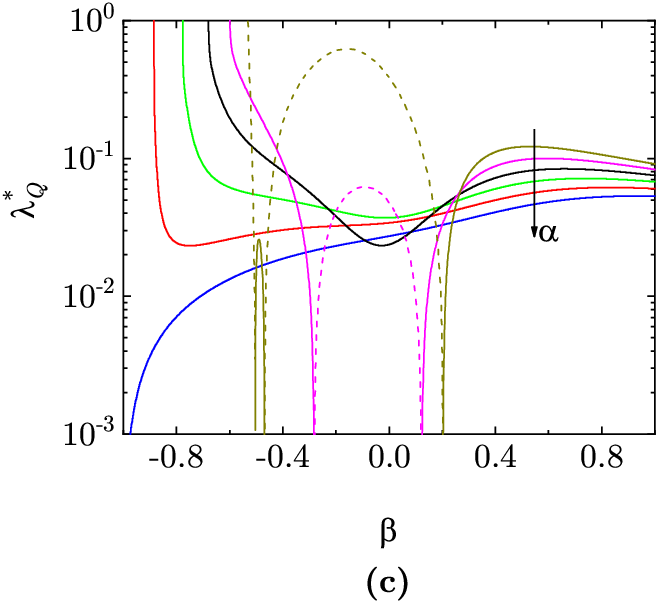}
\includegraphics[height=0.45\columnwidth]{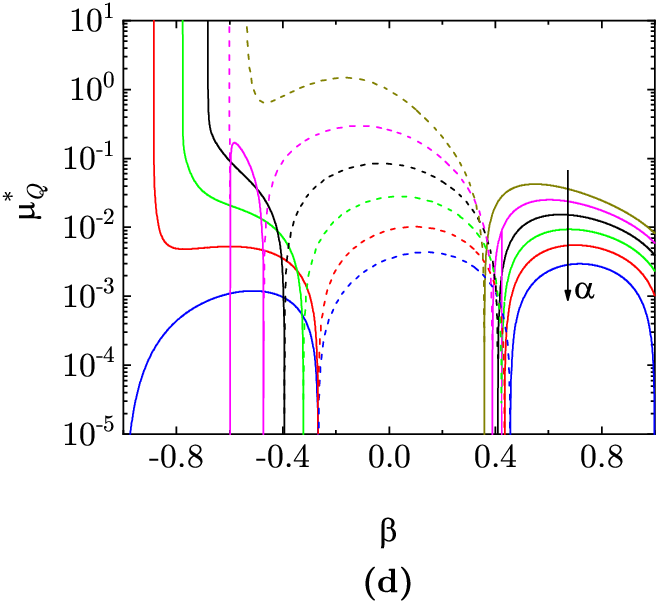}
  \caption{Plot of \textbf{a} the reduced thermal heat conductivity $\lambda^*$, \textbf{b} the reduced diffusive heat conductivity  $\mu^*$,  \textbf{c} the reduced thermal torque-vorticity conductivity $\lambda_Q^*$, and \textbf{d} the reduced diffusive torque-vorticity conductivity  $\mu_Q^*$  versus the coefficient of tangential restitution $\beta$ for $\alpha=0.5, 0.6,0.7,0.8,0.9,1$. The arrows indicate increasing values of $\alpha$.  Note the vertical asymptotes signaling the  threshold values $\beta_0(\alpha)$ below which the coefficients diverge. Note also that, in panels \textbf{c} and \textbf{d},  absolute values are taken and the dashed lines refer  to negative values. Here, the particles are assumed to have a uniform mass distribution ($\kappa=\frac{2}{5})$.}
  \label{fig2}
\end{figure}

\subsection{Coefficients $\lambda$, $\mu$, $\lambda_Q$, and $\mu_Q$}
Now we turn to the coefficients related to the heat flux and the torque-vorticity vector, as displayed in Fig.~\ref{fig2}.
Except for $\alpha=1$, the shapes of the curves of $\lambda^*$ and $\mu^*$ (see Figs.~\ref{fig2}a and \ref{fig2}b) differ qualitatively from those obtained from the IRHSM with the standard Sonine approximation because of the divergence of the IRMM coefficients at $\beta=\beta_0(\alpha)$. As said before, such a divergence is induced by that of the HCS fourth-degree cumulants \cite{KS22}.
This implies a breakdown of a hydrodynamic description, in analogy with what happens with the IMM \cite{S03,BGM10,KGS14}.

Figures~\ref{fig2}c and \ref{fig2}d show that the transport coefficients associated with the torque-vorticity vector $\mathbf{Q}$ (which is ignored in the standard Sonine approximation of the IRHSM) have a dependence on $\alpha$ and $\beta$ much more intricate than the heat-flux coefficients. In particular, both $\lambda_Q^*$ and $\nu_Q^*$ may attain negative values for intermediate and small values of $\beta$. This effect takes place even at $\alpha=1$, in which case $\lambda^*$ and $\mu^*$ remain finite for all $\alpha$.

\begin{figure}
\centering
\includegraphics[height=0.45\columnwidth]{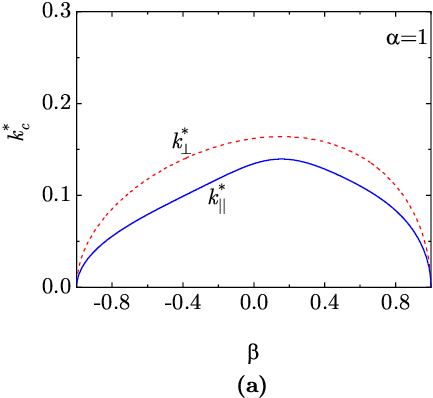}
\includegraphics[height=0.45\columnwidth]{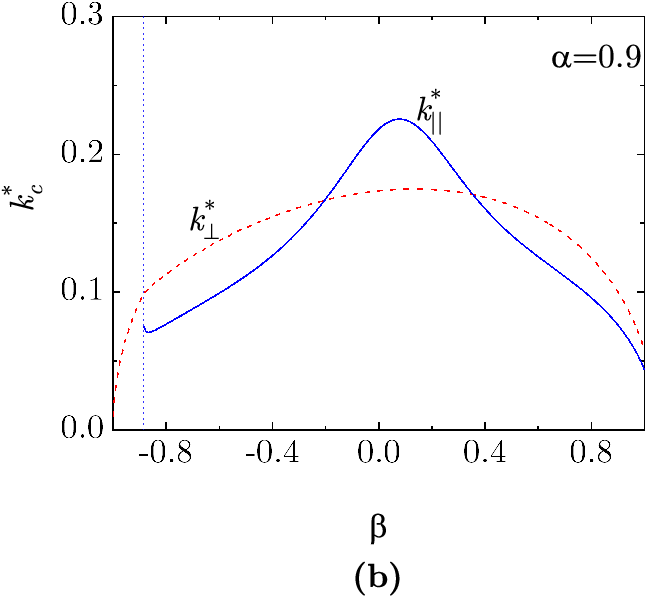}\\
\vspace{0.04cm}
\includegraphics[height=0.45\columnwidth]{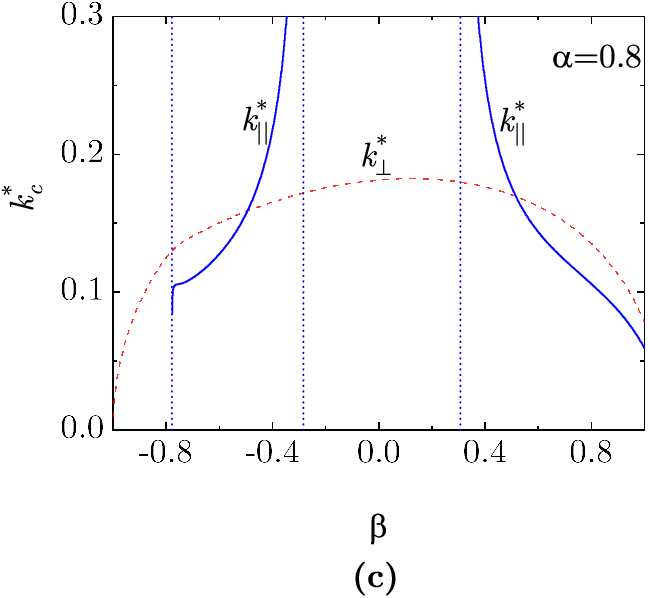}
\includegraphics[height=0.45\columnwidth]{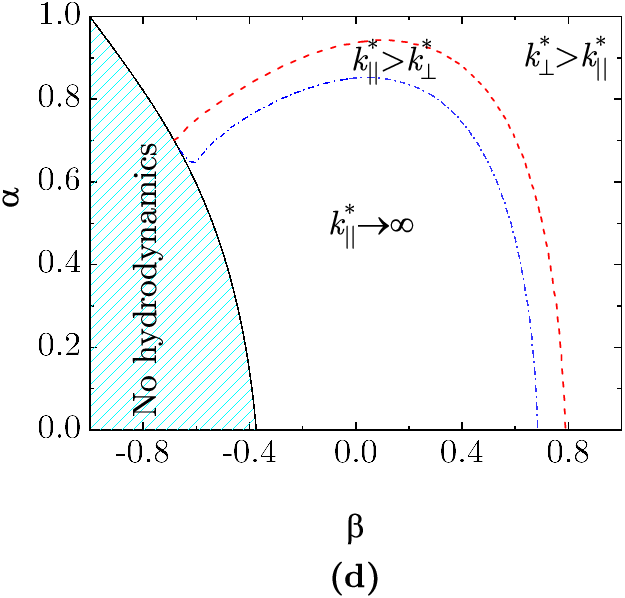}
  \caption{\textbf{a}--\textbf{c} Plot of $k_\perp^*$ (dashed lines) and $k_{\|}^*$ (solid lines)   versus the coefficient of tangential restitution $\beta$ for \textbf{a} $\alpha=1$, \textbf{b} $\alpha=0.9$, and \textbf{c} $\alpha=0.8$. The vertical dotted lines in panels \textbf{b} and \textbf{c} signal the divergence of  $k_{\|}^*$.  \textbf{d} Loci $\beta=\beta_0(\alpha)$ (solid line), $k_\perp^*=k_{\|}^*$ (dashed line), and $\lambda^*=\mu^*$ (dashed-dotted line) in the plane $\alpha$ versus $\beta$. Here, the particles are assumed to have a uniform mass distribution ($\kappa=\frac{2}{5})$.}
  \label{fig3}
\end{figure}

\subsection{Instability of the Homogeneous Cooling State}
Depending on the values of $\alpha$, $\beta$, and $\ka$, the HCS can be unstable versus long-wavelength perturbations. A standard linear stability analysis of the NSF hydrodynamic equations for the IRHSM \cite{GSK18,MS21b} can be straightforwardly extended to the IRMM. According to it, the critical wave number below which the HCS becomes unstable is $k_c=\max\{k_\perp,k_{\|}\}$, where
\beq
k_\perp=\sqrt{\frac{nm\zeta^{(0)}}{2\eta}},\quad k_{\|}=\sqrt{\frac{3n\zeta^{(0)}}{2(\lambda-\mu n/T)}}.
\eeq
While $k<k_\perp$ signals the appearance of vortices,  a clustering phenomenon is present if $k<k_{\|}$. The wave number $k_{\|}$ is not well defined if $\beta<\beta_0(\alpha)$ since both $\lambda$ and $\mu$ diverge in that region because of the divergence of the HCS cumulants.

Figures~\ref{fig3}a, \ref{fig3}b, and \ref{fig3}c show $k_\perp^*$ and $k_{\|}^*$, where $k^*=k\sqrt{\tau_t T/m}/\nuM^{(0)}$, versus $\beta$ for $\alpha=1$, $0.9$, and $0.8$, respectively. We observe that, in the case $\alpha=1$, $k_c=k_\perp$, whereas $k_c=k_{\|}$ in the interval $-0.203<\beta<0.355$ if $\alpha=0.9$. This behavior is qualitatively similar to that found in the IRHSM within the standard Sonine approximation. On the other hand, in the case $\alpha=0.8$, not only  $k_c=k_{\|}$ in the interval $-0.494<\beta<0.517$, but $k_c\to\infty$ in the inner interval $-0.283<\beta<0.308$. This divergence reflects the fact that $\lambda<\mu n/T$ in that region, implying the absolute instability of the HCS for any perturbation.
It turns out that the IRHSM with the standard Sonine approximation also predicts a region of absolute instability but confined to a much smaller region \cite{MS21b}.

Figure~\ref{fig3}d display the loci $\beta=\beta_0(\alpha)$, $k_{\perp}^*=k_{\|}^*$, and $\lambda^*=\mu^*$  in the plane $\alpha$ versus $\beta$. In the region below the locus  $\beta=\beta_0(\alpha)$, the HCS cumulants and the NSF heat-flux coefficients diverge; in the region below the locus $k_\perp^*=k_{\|}^*$, the critical wavenumber is the longitudinal one (i.e., $k_c^*=k_{\|}^*$); finally,
in the region below the locus $\lambda^*=\mu^*$, $k_c^*\to\infty$, implying the absolute instability of the HCS.

\begin{figure}
\centering
\includegraphics[height=0.44\columnwidth]{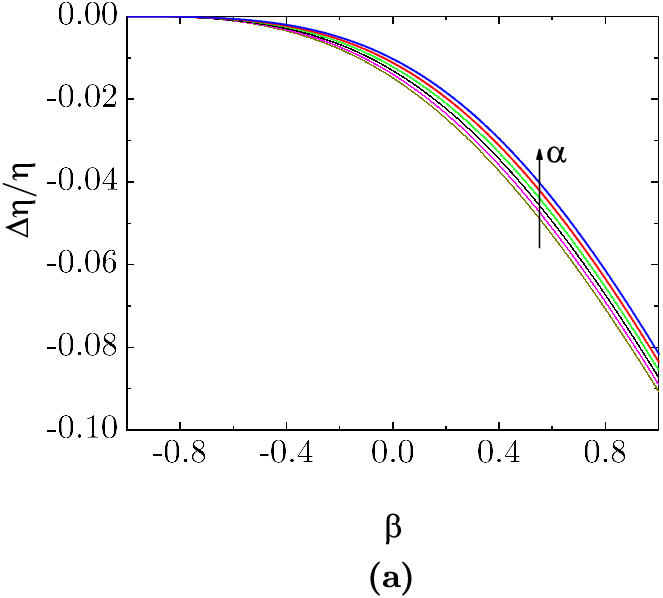}\\
\includegraphics[height=0.44\columnwidth]{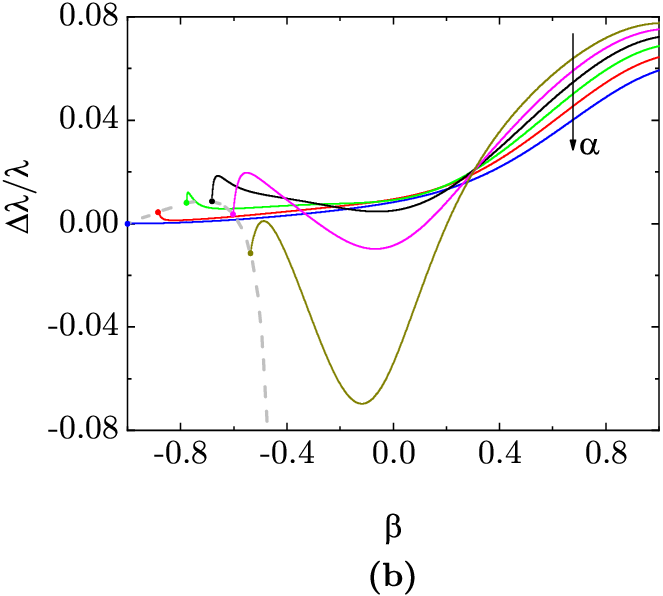}
\includegraphics[height=0.44\columnwidth]{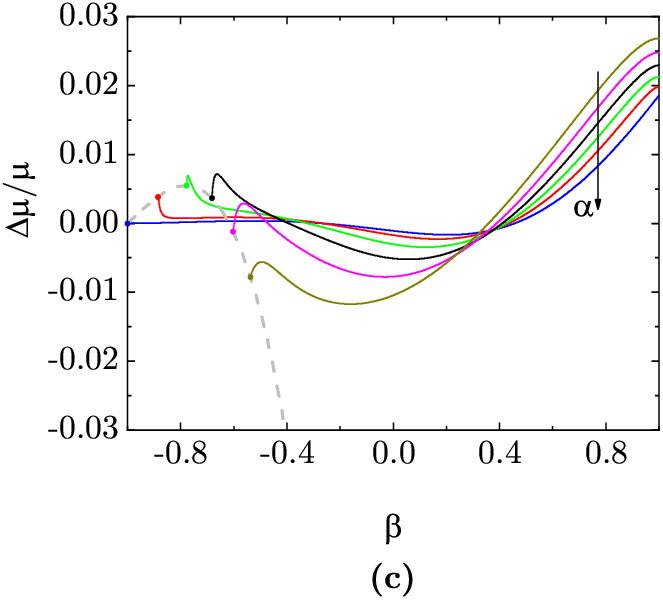}
  \caption{Plot of the relative deviations \textbf{a} $\Delta\eta/\eta$, \textbf{b} $\Delta \lambda/\lambda$, and \textbf{c} $\Delta \mu/\mu$ versus the coefficient of tangential restitution $\beta$ for $\alpha=0.5, 0.6,0.7,0.8,0.9,1$. The arrows indicate increasing values of $\alpha$. The  dashed lines in panels \textbf{b} and \textbf{c} represent the values of $\Delta \lambda/\lambda$ and $\Delta \mu/\mu$, respectively, at $\beta=\beta_0(\alpha)$. Here, the particles are assumed to have a uniform mass distribution ($\kappa=\frac{2}{5})$.}
  \label{fig4}
\end{figure}

\subsection{Impact of the couplings $\mathsf{\Pi}\leftrightarrow\mathsf{\Omega}$ and $\mathbf{q}\leftrightarrow\mathbf{Q}$}

As said before, one of the strong points of the IRMM is that it unveils the inherent couplings $\mathsf{\Pi}\leftrightarrow\mathsf{\Omega}$ and $\mathbf{q}\leftrightarrow\mathbf{Q}$. It is then in order to assess the impact of those couplings on the NSF transport coefficients $\eta$, $\lambda$, and $\mu$. On the other hand, the coefficients $\eta_b$ and $\xi$ are not affected by the couplings.

In the case of the shear viscosity, if the coupling $\mathsf{\Pi}\leftrightarrow\mathsf{\Omega}$ were ignored, then one would forget the term $\psi_{20\mid  02}\Omega_{ij}^{(1)}$ in the first row of Eq.~\eqref{22ab}. This is equivalent to formally setting $\psi_{20\mid  02}\to 0$ in Eq.~\eqref{eta}, with the result
\beq
\label{etatilde}
\widetilde{\eta}=\frac{n\tau_t T}{\nuM^{(0)}}\left[\psi_{20\mid  20}-\frac{1}{2}\frac{\zeta^{(0)}}{\nuM^{(0)}}\right]^{-1},
\eeq
where we have used a tilde to distinguish the approximate shear viscosity ($\widetilde{\eta}$) from the true one ($\eta$).

In the case of the heat-flux coefficients, ignorance of the coupling  is equivalent to formally setting $\mathbf{Q}^{(1)}\to 0$ in the first two rows of Eq.~\eqref{25}. Thus, instead of Eqs.~\eqref{37&38}, one would have
\begin{subequations}
\label{37&38tilde}
\beq
  \label{37tilde}
  \begin{bmatrix}
   \tau_t\widetilde{\lambda}_t\\
   \tau_r\widetilde{\lambda}_r
   \end{bmatrix}=\frac{nT\tau_t}{2m}\left(\nuM^{(0)}\widetilde{\mathsf{\Phi}}-2\zeta^{(0)}\widetilde{\mathsf{I}}\right)^{-1}
 \cdot
  \begin{bmatrix}
  5\tau_t\left(1+2a_{20}\right)\\
3\tau_r\left(1+2a_{11}\right)\\
 \end{bmatrix},
 \eeq
\beq
  \label{38tilde}
 \begin{bmatrix}
   \widetilde{\mu}_t\\
   \widetilde{\mu}_r
 \end{bmatrix}= \frac{T}{n}\left(\nuM^{(0)}\widetilde{\mathsf{\Phi}}-\frac{3}{2}\zeta^{(0)}\widetilde{\mathsf{I}}\right)^{-1}
  \cdot
 \left(
 \frac{n\tau_t T}{2m}
 \begin{bmatrix}
 5\tau_t a_{20}\\
3\tau_r a_{11}
 \end{bmatrix}
  +\zeta^{(0)}
 \begin{bmatrix}
   \tau_t\lambda_t\\
   \tau_r\lambda_r
 \end{bmatrix}
 \right),
 \eeq
\end{subequations}
where $\widetilde{\mathsf{I}}$ is the $2\times2$ unit matrix and
\beq
\widetilde{\mathsf{\Phi}}\equiv\begin{bmatrix}
   \varphi_{30\mid  30}&\frac{8}{\kappa}\varphi_{30\mid  12}\\
 \frac{\kappa}{4} \varphi_{12\mid  30}&  \varphi_{12\mid  12}^{(1)}
 \end{bmatrix}.
 \eeq

The relative deviations $\Delta \eta/\eta$, $\Delta \lambda/\lambda$, and $\Delta \mu/\mu$,  (where $\Delta \eta=\widetilde{\eta}-\eta$, $\Delta \lambda=\widetilde{\lambda}-\lambda$, and $\Delta \mu=\widetilde{\mu}-\mu$) are plotted in Fig.~\ref{fig4}.
As can be seen, the approximate shear viscosity $\tilde{\eta}$ underestimates the true value of $\eta$, this effect increasing monotonically with increasing inelasticity and, especially, increasing roughness. At $\beta=1$, the deviation of $\tilde{\eta}$ from $\eta$ ranges from about $9$\% for
$\alpha=0.5$ to about $8$\% for $\alpha=1$ (Pidduck's limit).

In what concerns $\Delta \lambda/\lambda$ and $\Delta \mu/\mu$, the combined influence of $\alpha$ and $\beta$ is much more intricate than in the case of $\Delta \eta/\eta$. In the region of high roughness (say $\beta>0.4$), both transport coefficients are overestimated if $\mathbf{Q}$ is ignored, this effect increasing again monotonically with increasing inelasticity and roughness. However, at intermediate roughness (say $-0.4<\beta<0.4$), $\tilde{\lambda}$ and $\tilde{\mu}$ typically underestimate the true values. As the threshold values $\beta_0(\alpha)$ are approached, the relative deviations  $\Delta \lambda/\lambda$ and $\Delta \mu/\mu$ reach finite values. At $\beta=1$, the deviations of the pair $(\tilde{\lambda},\tilde{\mu})$  range from about ($8$\%,$3$\%) for
$\alpha=0.5$ to about ($6$\%,$2$\%) for $\alpha=1$ (Pidduck's limit).
Note that, although at $\alpha=1$ both $\mu$ and $\widetilde{\mu}$ tend to $0$ in the limit $\beta\to 1$, the relative deviation $\Delta \mu/\mu$ tends to a finite value in that limit.

Except for extremely low values of $\alpha$, the consequences of disregarding the couplings $\mathsf{\Pi}\leftrightarrow\mathsf{\Omega}$ and $\mathbf{q}\leftrightarrow\mathbf{Q}$ are generally not substantial. However, to maintain consistency, these couplings must be considered.

\subsection{Proposal of an augmented Sonine approximation for the IRHSM}
As said before, the derivation of the NSF transport coefficients in the IRHSM requires the use of approximations, since taking velocity moments in the kinetic equations for $f^{(0)}$ and $f^{(1)}$ generates an infinite hierarchy of moment equations. In the standard Sonine approximation \cite{KSG14,MS21a}, the unknown vector functions $\bbA$ and $\bbB$ appearing in Eq.~\eqref{f1} are approximated by a Maxwellian times a linear combination of $\mathbf{V}$, $V^2\mathbf{V}$, and  $\omega^2\mathbf{V}$. Likewise, the unknown tensor function $\bbC_{ij}$ is approximated by a Maxwellian times $V_iV_j-\frac{1}{3}V^2\delta_{ij}$.
In the light of the results we have obtained in this paper for the IRMM, we propose here for the IRHSM  the more consistent Sonine approximation
\begin{subequations}
\label{augmSA}
\bal
\bbA\to & -f_M \left\{\gamma_{A_t}\left(V^2-\frac{5\tau_t T}{m}\right)\mathbf{V}+\gamma_{A_r}\left(\omega^2-\frac{3\tau_r T}{I}\right)\mathbf{V}
\right.\nn&\left.
+\gamma_{A_Q}\left[(\mathbf{V}\cdot\bw)\bw-\frac{\tau_r T}{I}\bV\right]\right\},
\label{Aapprox}
\eal
\bal
\bbB\to & -f_M \left\{\gamma_{B_t}\left(V^2-\frac{5\tau_t T}{m}\right)\mathbf{V}+\gamma_{B_r}\left(\omega^2-\frac{3\tau_r T}{I}\right)\mathbf{V}\right.\nn
&\left.+\gamma_{B_Q}\left[(\mathbf{V}\cdot\bw)\bw-\frac{\tau_r T}{I}\bV\right]\right\},
\label{Bapprox}
\eal
\beq
\bbC_{ij}\to -f_M\left[{\gamma_{C_t}} \left(V_iV_j-\frac{1}{3}V^2\delta_{ij}\right)+{\gamma_{C_r}} \left(\omega_i\omega_j-\frac{1}{3}\omega^2\delta_{ij}\right)\right],
\eeq
\end{subequations}
where
\beq
f_M=n\left(\frac{m I}{4\pi^2\tau_t\tau_r T^2}\right)^{\frac{3}{2}}\exp\left(-\frac{mV^2}{2\tau_t T}-\frac{I\omega^2}{2\tau_r T}\right).
\label{Maxw}
\eeq
In the standard Sonine approximation \cite{KSG14,MS21a}, one assumes $\gamma_{A_Q}=\gamma_{B_Q}=\gamma_{C_r}=0$. On the other hand, in the augmented version given by Eqs.~\eqref{augmSA}, the coefficient $\gamma_{C_r}$ couples $\mathsf{\Pi}$ and $\mathsf{\Omega}$, whereas the coefficients $\gamma_{A_Q}$ and $\gamma_{B_Q}$ couple $\mathbf{q}_t$ and $\mathbf{q}_r$ to $\mathbf{Q}$.

It is worthwhile noting that, in the case of hard disks on a plane \cite{MS21a}, one has $\mathbf{V}\perp\bw$ and $\bw\|\widehat{\mathbf{z}}$, so that
$(\mathbf{V}\cdot\bw)\bw=0$ and $\omega_i\omega_j-\frac{1}{3}\omega^2\delta_{ij}=0$. Therefore, in that case the augmented Sonine approximation becomes identical to the standard one.

\section{Conclusions}
\label{sec5}

Quoting Ernst and Brito \cite{EB02a,EB02b}, one can say that ``what harmonic oscillators are for quantum
mechanics, and dumb-bells for polymer physics, that is what elastic and
inelastic Maxwell models are for kinetic theory.'' The present paper, in conjunction with Ref.~\cite{KS22}, aims to extend Ernst and Brito's dictum into the domain of models featuring rough particles, specifically the IRMM.

Building upon the exact collisional production rates established in Ref.~\cite{KS22}, in this work we have derived the NSF transport coefficients as explicit functions of the coefficients of normal ($\alpha$) and tangential ($\beta$) restitution, along with the reduced moment of inertia ($\kappa$). The resulting hydrodynamic constitutive equations, Eqs.~\eqref{NSF}, encompass the shear viscosity ($\eta$), the bulk viscosity ($\eta_b$), the thermal heat conductivity  ($\lambda$), the diffusive heat conductivity ($\mu$), and the cooling-rate transport coefficient ($\xi$). The evaluation of $\eta$ requires the parallel evaluation of the spin viscosity $\eta_\Omega$ [see Eq.~\eqref{Piij1}], whereas the evaluation of $\lambda$ and $\mu$ requires the parallel evaluation of the torque-vorticity coefficients $\lambda_Q$ and $\mu_Q$ [see Eq.~\eqref{lambdaQmuQ}].
Interestingly, Figs.~\ref{fig1} and \ref{fig2} illustrate the complex dependence of these eight transport coefficients on $\alpha$ and $\beta$ for a uniform mass distribution  ($\kappa=\frac{2}{5}$).

Our analysis reveals a noteworthy result as both the mean spin vector and the couple stress tensor vanish to the NSF order, contrary to the case of micropolar fluids \cite{E66, B97, L99b, MHN02} and dense gases \cite{MSD66,DT75,GK91,K24}. This discrepancy is attributed to the low-density regime and the absence of boundary effects in the NSF description.

While the coefficients tied to pressure and spin-spin tensors remain finite for any $\alpha$ and $\beta$, those linked to the heat flux and the torque-vorticity vector exhibit dependency on the HCS cumulants defined in Eqs.~\eqref{cumu}, diverging if $\beta$ is less than an $\alpha$-dependent value, $\beta_0(\alpha)$ \cite{KS22}. This divergence, originating from an algebraic high-velocity tail in the HCS distribution function \cite{EB02a,EB02b,EB02c}, manifests in the breakdown of hydrodynamics if $\beta<\beta_0(\alpha)$. Beyond this region, the NSF hydrodynamic framework aids in determining the instability of the HCS against weak inhomogeneous perturbations.
That instability occurs if the wave number of the perturbations is smaller than $k_c=\max\{k_\perp,k_{\|}\}$ (see Fig.~\ref{fig3}). Remarkably, a dome-shaped region emerges in the $\alpha$ vs $\beta$ plane, within which $k_c\to\infty$, signifying an absolute instability of the HCS.

Additionally, considering the couplings $\mathsf{\Pi}\leftrightarrow \mathsf{\Omega}$ and $\mathbf{q}\leftrightarrow \mathbf{Q}$, their impact on the NSF coefficients is found to be below 10\% if $\alpha>0.5$ and $\kappa=\frac{2}{5}$ (see Fig.~\ref{fig4}). Nevertheless, a consistent treatment should incorporate these couplings, even within the IRHSM.

Although, for simplicity and conciseness, this paper presents graphs exclusively for a uniform mass distribution ($\kappa=\frac{2}{5}$), the reduced moment of inertia $\kappa$ serves as an additional control parameter influencing results. The nontrivial impact of $\kappa$  on the final outcomes can be exemplified by the Pidduck limit ($\alpha=\beta=1$). According to the fourth column of Table \ref{table2}, the reduced shear viscosity, bulk viscosity, and thermal heat conductivity  take the values $(\eta^*,\eta_b^*,\lambda^*)=(1,\infty ,1.31)$, $(0.92,0.61 ,1.23)$, $(0.91,0.25 ,1.29)$, and $(0.98,0.21 ,1.30)$ for $\kappa=0$, $\frac{1}{10}$,  $\frac{2}{5}$, and $\frac{2}{3}$, respectively.

In summary, the IRMM serves as a compelling mathematical model, providing exact results and offering insights into the intricacies anticipated in the more realistic IRHSM. Future work aims to explore the exact non-Newtonian properties of the IRMM under simple shear flow.

\begin{appendices}

\section{Explicit expressions for the coefficients appearing in Eqs.~\eqref{6A-6H}}
\label{appB}
\setcounter{equation}{0}

As explained in Ref.~\cite{KS22}, we have adopted a specific criterion for coefficient notation. Firstly, assume that $\Psi_{k_1k_2}(\bxi)$ represents a homogeneous velocity polynomial with degrees $k_1$ and $k_2$ relative to $\bV$ and $\bw$, respectively. Thus, a coefficient such as $Y_{k_1 k_2\mid \ell_1\ell_2}$ (with $Y=\chi$, $\varphi$, or $\psi$) corresponds to the collisional moment $\mathcal{J}[\Psi_{k_1k_2}]$, linked to a product like $\langle \Psi_{i_1i_2}(\bxi)\rangle\langle \Psi_{j_1j_2}(\bxi)\rangle$ with $i_1+j_1=\ell_1$ and $i_2+j_2=\ell_2$. If there exists more than one $Y_{k_1 k_2\mid \ell_1\ell_2}$ for a given $\mathcal{J}[\Psi_{k_1k_2}]$, each is identified by a superscript. Lastly, Greek letters $\chi$, $\varphi$, and $\psi$ signify coefficients in $\mathcal{J}[\Psi_{k_1k_2}]$ linked to scalar, vector, and tensor quantities $\Psi_{k_1k_2}(\bxi)$, respectively; additionally, an overline is used if $\Psi_{k_1k_2}(\bxi)$ involves the inner product $\bV\cdot\bw$.

The $17$ coefficients appearing in Eqs.~\eqref{6A-6H} are given by
\begin{subequations}
\bal
\label{B1}
\varphi_{01\mid 01}=&\frac{4\bt}{3\ka},\quad \psi_{11\mid 11}=\frac{1}{3}\left(\at+2\bt\frac{1+\ka}{\ka}\right),\\
\label{B2}
\chi_{20\mid 20}=&\frac{2}{3}\left[\at\left(1-\at\right)+2\bt\left(1-\bt\right)\right],\quad \chi_{20\mid 02}=-\frac{\bt^2}{3},\\
\label{B4}
\chi_{02\mid 02}=&\frac{4\bt}{3\ka}\left(1-\frac{\bt}{\ka}\right),\quad \chi_{02\mid 20}=-\frac{16\bt^2}{3\ka^2},\\
\label{B3}
\psi_{20\mid 20}=&\frac{2}{15}\left(5\at-2\at^2-6\at\bt+10\bt-7\bt^2\right),\\
\psi_{20\mid 02}=&\frac{\bt^2}{6}, \quad \psi_{02\mid 20}=\frac{8\bt^2}{3\ka^2},\quad \psi_{02\mid 02}=\frac{2\bt}{15\ka}\left(10-\frac{7\bt}{\ka}\right),
 \\
\varphi_{30\mid 30}=&\frac{1}{15}\left(15\at-11\at^2-8\at\bt+30\bt-26\bt^2\right),\\
\varphi_{30\mid 12}=&-\frac{\bt^2}{6},\quad \varphi_{12\mid 30}=-\frac{8\bt^2}{3\ka^2}\\
\varphi_{12\mid 12}^{(1)}=&\frac{1}{15}\left[5\at+10\bt+\frac{2\bt}{\ka}\left(10-4\at-11\bt-\frac{5\bt}{\ka}\right)\right]
,\\
\overline{\varphi}_{12\mid 12}^{(1)}=&
\frac{1}{15}\left[5\at+10\bt+\frac{\bt}{\ka}\left(20-3\at-7\bt\frac{1+\ka}{\ka}\right)\right]
,\\
\varphi_{12\mid 12}^{(2)}=&\frac{2\bt}{15\ka}\left(2\at+3\bt\right)
,\quad \overline{\varphi}_{12\mid 12}^{(2)}=\frac{\bt}{15\ka}\left(\at-\bt\frac{1+\ka}{\ka}\right),
\eal
\end{subequations}
where
\beq
\label{7}
 \widetilde\alpha\equiv\frac{1+\alpha}{2},\quad \widetilde\beta\equiv\frac{1+\beta}{2}\frac{\kappa}{1+\kappa}.
 \eeq

\end{appendices}

\backmatter

\bmhead{Acknowledgments}

A.S.\ acknowledges financial support from Grant No.~PID2020-112936GB-I00 funded by MCIN/AEI/10.13039/501100011033, and from Grant No.~IB20079 funded by Junta de
Extremadura (Spain) and by ``ERDF A way of making Europe.''
G.M.K.\ is grateful to the Conselho Nacional de Desenvolvimento Cient\'{i}fico e Tecnol\'{o}gico (CNPq) for financial support through Grant No.\  304054/2019-4.

\bmhead{Data availability}
The datasets employed to generate Figs.~\ref{fig1}--\ref{fig4}  are available from the corresponding author on reasonable request.

\section*{Declarations}
\bmhead{Conflict of interest}
The authors declare no conflicts of interest that are relevant to the content of this article.






    \bibliography{C:/AA_D/Dropbox/Mis_Dropcumentos/bib_files/Granular}

\begin{thebibliography}{10}
\providecommand{\url}[1]{{#1}}
\providecommand{\urlprefix}{URL }
\providecommand{\doi}[1]{\url{https://doi.org/#1}}
\bibcommenthead

\bibitem{D01}
J.W. Dufty, Kinetic theory and hydrodynamics for a low density granular gas.
\newblock Adv. Complex Syst. \textbf{4}, 397--406 (2001).
\newblock \doi{10.1142/S0219525901000395}

\bibitem{BP04}
N.V. Brilliantov, T.~P\"oschel, \emph{Kinetic Theory of Granular Gases} (Oxford
  University Press, Oxford, 2004)

\bibitem{G19}
V.~Garz\'o, \emph{Granular Gaseous Flows. A Kinetic Theory Approach to Granular
  Gaseous Flows} (Springer Nature, Switzerland, 2019)

\bibitem{C90}
C.S. Campbell, Rapid granular flows.
\newblock Annu. Rev. Fluid Mech. \textbf{22}, 57--92 (1990).
\newblock \doi{10.1146/annurev.fl.22.010190.000421}

\bibitem{BDKS98}
J.J. Brey, J.W. Dufty, C.S. Kim, A.~Santos, Hydrodynamics for granular flow at
  low density.
\newblock Phys. Rev. E \textbf{58}, 4638--4653 (1998).
\newblock \doi{10.1103/PhysRevE.58.4638}

\bibitem{GDH07}
V.~Garz\'o, J.W. Dufty, C.M. Hrenya, Enskog theory for polydisperse granular
  mixtures. {I. Navier-Stokes} order transport.
\newblock Phys. Rev. E \textbf{76}, {031}{303} (2007).
\newblock \doi{10.1103/PhysRevE.76.031303}

\bibitem{GSM07}
V.~Garz\'o, A.~Santos, J.M. Montanero, Modified {Sonine} approximation for the
  {Navier--Stokes} transport coefficients of a granular gas.
\newblock Physica A \textbf{376}, 94--107 (2007).
\newblock \doi{10.1016/j.physa.2006.10.081}

\bibitem{JR85a}
J.T. Jenkins, M.W. Richman, Kinetic theory for plane flows of a dense gas of
  identical, rough, inelastic, circular disks.
\newblock Phys. Fluids \textbf{28}, 3485--3494 (1985).
\newblock \doi{10.1063/1.865302}

\bibitem{LS87}
C.K.K. Lun, S.B. Savage, A simple kinetic theory for granular flow of rough,
  inelastic, spherical particles.
\newblock J. Appl. Mech. \textbf{54}, 47--53 (1987).
\newblock \doi{10.1115/1.3172993}

\bibitem{ZTPSH98}
P.~Zamankhan, H.V. Tafreshi, W.~Polashenski, P.~Sarkomaa, C.L. Hyndman, Shear
  induced diffusive mixing in simulations of dense {Couette} flow of rough,
  inelastic hard spheres.
\newblock J. Chem. Phys. \textbf{109}, 4487--4491 (1998).
\newblock \doi{10.1063/1.477076}

\bibitem{LHMZ98}
S.~Luding, M.~Huthmann, S.~McNamara, A.~Zippelius, Homogeneous cooling of
  rough, dissipative particles: Theory and simulations.
\newblock Phys. Rev. E \textbf{58}, 3416--3425 (1998).
\newblock \doi{10.1103/PhysRevE.58.3416}

\bibitem{CLH02}
R.~Cafiero, S.~Luding, H.J. Herrmann, Rotationally driven gas of inelastic
  rough spheres.
\newblock Europhys. Lett. \textbf{60}, 854--860 (2002).
\newblock \doi{10.1209/epl/i2002-00295-7}

\bibitem{BPKZ07}
N.V. Brilliantov, T.~P\"oschel, W.T. Kranz, A.~Zippelius, Translations and
  rotations are correlated in granular gases.
\newblock Phys. Rev. Lett. \textbf{98}, {128}{001} (2007).
\newblock \doi{10.1103/PhysRevLett.98.128001}

\bibitem{CP08}
F.~Cornu, J.~Piasecki, Granular rough sphere in a low-density thermal bath.
\newblock Physica A \textbf{387}, 4856--4862 (2008).
\newblock \doi{10.1016/j.physa.2008.03.014}

\bibitem{SKG10}
A.~Santos, G.M. Kremer, V.~Garz\'o, Energy production rates in fluid mixtures
  of inelastic rough hard spheres.
\newblock Prog. Theor. Phys. Suppl. \textbf{184}, 31--48 (2010).
\newblock \doi{10.1143/PTPS.184.31}

\bibitem{GG20}
R.~{G\'omez Gonz\'alez}, V.~Garz\'o, Non-{N}ewtonian rheology in inertial
  suspensions of inelastic rough hard spheres under simple shear flow.
\newblock Phys. Fluids \textbf{32}, {073}{315} (2020).
\newblock \doi{10.1063/5.0015241}

\bibitem{KSG14}
G.M. Kremer, A.~Santos, V.~Garz\'o, Transport coefficients of a granular gas of
  inelastic rough hard spheres.
\newblock Phys. Rev. E \textbf{90}, {022}{205} (2014).
\newblock \doi{10.1103/PhysRevE.90.022205}

\bibitem{GSK18}
V.~Garz\'o, A.~Santos, G.M. Kremer, Impact of roughness on the instability of a
  free-cooling granular gas.
\newblock Phys. Rev. E \textbf{97}, {052}{901} (2018).
\newblock \doi{10.1103/PhysRevE.97.052901}

\bibitem{MS21a}
A.~Meg\'ias, A.~Santos, Hydrodynamics of granular gases of inelastic and rough
  hard disks or spheres. {I}. {T}ransport coefficients.
\newblock Phys. Rev. E \textbf{104}, {034}{901} (2021).
\newblock \doi{10.1103/PhysRevE.104.034901}

\bibitem{MS21b}
A.~Meg\'ias, A.~Santos, Hydrodynamics of granular gases of inelastic and rough
  hard disks or spheres. {II}. {S}tability analysis.
\newblock Phys. Rev. E \textbf{104}, {034}{902} (2021).
\newblock \doi{10.1103/PhysRevE.104.034902}

\bibitem{M67}
J.C. Maxwell, {IV}. {O}n the dynamical theory of gases.
\newblock Phil. Trans. Roy. Soc. (London) \textbf{157}, 49--88 (1867).
\newblock \doi{10.1098/rstl.1867.0004}

\bibitem{TM80}
C.~Truesdell, R.G. Muncaster, \emph{{Fundamentals of Maxwell's Kinetic Theory
  of a Simple Monatomic Gas}} (Academic Press, New York, 1980)

\bibitem{GS03}
V.~Garz{\'o}, A.~Santos, \emph{Kinetic Theory of Gases in Shear Flows:
  Nonlinear Transport}.
\newblock Fundamental Theories of Physics (Springer, Dordrecht, 2003)

\bibitem{S09b}
A.~Santos, Solutions of the moment hierarchy in the kinetic theory of {Maxwell}
  models.
\newblock Cont. Mech. Thermodyn. \textbf{21}, 361--387 (2009).
\newblock \doi{10.1007/s00161-009-0113-5}

\bibitem{E81}
M.H. Ernst, Nonlinear model-{B}oltzmann equations and exact solutions.
\newblock Phys. Rep. \textbf{78}, 1--171 (1981).
\newblock \doi{10.1016/0370-1573(81)90002-8}

\bibitem{BCG00}
A.V. Bobylev, J.A. Carrillo, I.M. Gamba, On some properties of kinetic and
  hydrodynamic equations for inelastic interactions.
\newblock J. Stat. Phys. \textbf{98}, 743--773 (2000).
\newblock \doi{10.1023/A:1018627625800}

\bibitem{CCG00}
J.A. Carrillo, C.~Cercignani, I.M. Gamba, Steady states of a {Boltzmann}
  equation for driven granular media.
\newblock Phys. Rev. E \textbf{62}, 7700--7707 (2000).
\newblock \doi{10.1103/PhysRevE.62.7700}

\bibitem{BK00}
E.~Ben-Naim, P.L. Krapivsky, Multiscaling in inelastic collisions.
\newblock Phys. Rev. E \textbf{61}, R5--R8 (2000).
\newblock \doi{10.1103/PhysRevE.61.R5}

\bibitem{C01b}
C.~Cercignani, Shear flow of a granular material.
\newblock J. Stat. Phys. \textbf{102}, 1407--1415 (2001).
\newblock \doi{10.1023/A:1004804815471}

\bibitem{BC02}
A.V. Bobylev, C.~Cercignani, Moment equations for a granular material in a
  thermal bath.
\newblock J. Stat. Phys. \textbf{106}, 547--567 (2002).
\newblock \doi{10.1023/A:1013754205008}

\bibitem{KB02}
P.L. Krapivsky, E.~Ben-Naim, Nontrivial velocity distributions in inelastic
  gases.
\newblock J. Phys. A: Math. Gen. \textbf{35}, L147--L152 (2002).
\newblock \doi{10.1088/0305-4470/35/11/103}

\bibitem{BK02}
E.~Ben-Naim, P.L. Krapivsky, Scaling, multiscaling, and nontrivial exponents in
  inelastic collision processes.
\newblock Phys. Rev. E \textbf{66}, {011}{309} (2002).
\newblock \doi{10.1103/PhysRevE.66.011309}

\bibitem{BK02b}
E.~Ben-Naim, P.L. Krapivsky, Impurity in a {M}axwellian unforced granular
  fluid.
\newblock Eur. Phys. J. E \textbf{8}, 507--515 (2002).
\newblock \doi{10.1140/epje/i2002-10034-0}

\bibitem{EB02a}
M.H. Ernst, R.~Brito, High-energy tails for inelastic {Maxwell} models.
\newblock Europhys. Lett. \textbf{58}, 182--187 (2002).
\newblock \doi{10.1209/epl/i2002-00622-0}

\bibitem{EB02b}
M.H. Ernst, R.~Brito, Scaling solutions of inelastic {Boltzmann} equations with
  over-populated high energy tails.
\newblock J. Stat. Phys. \textbf{109}, 407--432 (2002).
\newblock \doi{10.1023/A:1020437925931}

\bibitem{EB02c}
M.H. Ernst, R.~Brito, Driven inelastic {Maxwell} models with high energy tails.
\newblock Phys. Rev. E \textbf{65}, {040}{301}(R) (2002).
\newblock \doi{10.1103/PhysRevE.65.040301}

\bibitem{BBP02}
A.~Baldassarri, U.M. {Bettolo Marconi}, A.~Puglisi, Influence of correlations
  on the velocity statistics of scalar granular gases.
\newblock Europhys. Lett. \textbf{58}, 14--20 (2002).
\newblock \doi{10.1209/epl/i2002-00600-6}

\bibitem{BP02a}
U.M. {Bettolo Marconi}, A.~Puglisi, Mean-field model of free-cooling inelastic
  mixtures.
\newblock Phys. Rev. E \textbf{65}, 051,305 (2002).
\newblock \doi{10.1103/PhysRevE.65.051305}

\bibitem{BP02b}
U.M. {Bettolo Marconi}, A.~Puglisi, Steady-state properties of a mean-field
  model of driven inelastic mixtures.
\newblock Phys. Rev. E \textbf{66}, 011,301 (2002).
\newblock \doi{10.1103/PhysRevE.66.011301}

\bibitem{EB03}
M.H. Ernst, R.~Brito, in \emph{Granular Gas Dynamics}, \emph{Lecture Notes in
  Physics}, vol. 624, ed. by T.~P\"oschel, S.~Luding (Springer, Berlin, 2003),
  pp. 3--36.
\newblock \doi{10.1007/978-3-540-39843-1_1}

\bibitem{BK03}
E.~Ben-Naim, P.L. Krapivsky, in \emph{Granular Gas Dynamics}, \emph{Lecture
  Notes in Physics}, vol. 624, ed. by T.~P\"oschel, S.~Luding (Springer,
  Berlin, 2003), pp. 65--94.
\newblock \doi{10.1007/978-3-540-39843-1_3}

\bibitem{BC03}
A.V. Bobylev, C.~Cercignani, Self-similar asymptotics for the {Boltzmann}
  equation with inelastic and elastic interactions.
\newblock J. Stat. Phys. \textbf{110}, 333--375 (2003).
\newblock \doi{10.1023/A:1021031031038}

\bibitem{BCT03}
A.V. Bobylev, C.~Cercignani, G.~Toscani, Proof of an asymptotic property of
  self-similar solutions of the {Boltzmann} equation for granular materials.
\newblock J. Stat. Phys. \textbf{111}, 403--417 (2003).
\newblock \doi{10.1023/A:1022273528296}

\bibitem{SE03}
A.~Santos, M.H. Ernst, Exact steady-state solution of the {B}oltzmann equation:
  {A} driven one-dimensional inelastic {M}axwell gas.
\newblock Phys. Rev. E \textbf{68}, {011}{305} (2003).
\newblock \doi{10.1103/PhysRevE.68.011305}

\bibitem{BG06}
A.V. Bobylev, I.M. Gamba, Boltzmann equations for mixtures of {Maxwell} gases:
  exact solutions and power like tails.
\newblock J. Stat. Phys. \textbf{124}, 497--516 (2006).
\newblock \doi{10.1007/s10955-006-9044-8}

\bibitem{ETB06a}
M.H. Ernst, E.~Trizac, A.~Barrat, The rich behavior of the {B}oltzmann equation
  for dissipative gases.
\newblock Europhys. Lett. \textbf{76}, 56--62 (2006).
\newblock \doi{10.1209/epl/i2006-10225-3}

\bibitem{ETB06b}
M.H. Ernst, E.~Trizac, A.~Barrat, The {B}oltzmann equation for driven systems
  of inelastic soft spheres.
\newblock J. Stat. Phys. \textbf{124}, 549--586 (2006).
\newblock \doi{10.1007/s10955-006-9062-6}

\bibitem{BC07}
F.~Bolley, J.A. Carrillo, Tanaka theorem for inelastic {Maxwell} models.
\newblock Commun. Math. Phys. \textbf{276}, 287--314 (2007).
\newblock \doi{10.1007/s00220-007-0336-x}

\bibitem{GS07}
V.~Garz\'o, A.~Santos, Third and fourth degree collisional moments for
  inelastic {Maxwell} model.
\newblock J. Phys. A: Math. Theor. \textbf{40}, {14,927}--{14,943} (2007).
\newblock \doi{10.1088/1751-8113/40/50/002}

\bibitem{BTE07}
A.~Barrat, E.~Trizac, M.H. Ernst, Quasi-elastic solutions to the nonlinear
  {B}oltzmann equation for dissipative gases.
\newblock J. Phys. A: Math. Theor. \textbf{40}, 4057--4073 (2007).
\newblock \doi{10.1088/1751-8113/40/15/001}

\bibitem{GT12b}
V.~Garz\'o, E.~Trizac, Dissipative homogeneous {M}axwell mixtures: ordering
  transitionin the tracer limit.
\newblock Granul. Matter \textbf{14}, 99--104 (2012).
\newblock \doi{10.1007/s10035-011-0304-1}

\bibitem{SG12}
A.~Santos, V.~Garz\'o, Collisional rates for the inelastic {M}axwell model.
  {A}pplication to the divergence of anisotropic high-order velocity moments in
  the homogeneous cooling state.
\newblock Gran. Matt. \textbf{14}, 105--110 (2012).
\newblock \doi{10.1007/s10035-012-0336-1}

\bibitem{S03}
A.~Santos, Transport coefficients of $d$-dimensional inelastic {Maxwell}
  models.
\newblock Physica A \textbf{321}, 442--466 (2003).
\newblock \doi{10.1016/S0378-4371(02)01005-1}

\bibitem{GA05}
V.~Garz\'o, A.~Astillero, Transport coefficients for inelastic {M}axwell
  mixtures.
\newblock J. Stat. Phys. \textbf{118}, 935--971 (2005).
\newblock \doi{10.1007/s10955-004-2006-0}

\bibitem{CGV14}
M.G. Chamorro, V.~Garz\'o, F.~{Vega Reyes}, Navier--{S}tokes transport
  coefficients for driven inelastic {M}axwell models.
\newblock J. Stat. Mech. \textbf{P06008} (2014).
\newblock \doi{10.1088/1742-5468/2014/06/P06008}

\bibitem{GKT15}
V.~Garz\'o, N.~Khalil, E.~Trizac, Anomalous transport of impurities in
  inelastic {M}axwell gases.
\newblock Eur. Phys. J. E \textbf{38}, 16 (2015).
\newblock \doi{10.1140/epje/i2015-15016-5}

\bibitem{KG15}
A.~Kubicki, V.~Garz\'o, Inelastic {M}axwell models for monodisperse gas--solid
  flows.
\newblock J. Stat. Mech. \textbf{P03015} (2015).
\newblock \doi{10.1088/1742-5468/2015/03/P03015}

\bibitem{KG20}
N.~Khalil, V.~Garz\'o, Unified hydrodynamic description for driven and undriven
  inelastic {M}axwell mixtures at low density.
\newblock J. Phys. A: Math. Theor. \textbf{53}, {355}{002} (2020).
\newblock \doi{10.1088/1751-8121/ab9f72}

\bibitem{KGS14}
N.~Khalil, V.~Garz\'o, A.~Santos, Hydrodynamic {B}urnett equations for
  inelastic {M}axwell models of granular gases.
\newblock Phys. Rev. E \textbf{89}, {052}{201} (2014).
\newblock \doi{10.1103/PhysRevE.89.052201}

\bibitem{G03a}
V.~Garz\'o, Nonlinear transport in inelastic {M}axwell mixtures under simple
  shear flow.
\newblock J. Stat. Phys. \textbf{112}, 657--683 (2003).
\newblock \doi{10.1023/A:1023828109434}

\bibitem{SG07}
A.~Santos, V.~Garz\'o, Simple shear flow in inelastic {M}axwell models.
\newblock J. Stat. Mech. \textbf{P08021} (2007).
\newblock \doi{10.1088/1742-5468/2007/08/P08021}

\bibitem{G07}
V.~Garz\'o, Shear-rate dependent transport coefficients for inelastic {Maxwell}
  models.
\newblock J. Phys. A: Math. Theor. \textbf{40}, 10,729--10,767 (2007).
\newblock \doi{10.1088/1751-8113/40/35/002}

\bibitem{G08c}
V.~Garz\'o, Mass flux of a binary mixture of maxwell molecules under shear
  flow.
\newblock Physica A \textbf{387}, 3423--3431 (2008).
\newblock \doi{10.1016/j.physa.2008.02.019}

\bibitem{SGV09}
A.~Santos, V.~Garz\'o, F.~{Vega Reyes}, An exact solution of the inelastic
  {Boltzmann} equation for the {Couette} flow with uniform heat flux.
\newblock Eur. Phys. J. Spec. Top. \textbf{179}, 141--156 (2009).
\newblock \doi{10.1140/epjst/e2010-01199-9}

\bibitem{GT10}
V.~Garz\'o, E.~Trizac, Rheological properties for inelastic {M}axwell mixtures
  under shear flow.
\newblock J. Non-Newton. Fluid Mech. \textbf{165}, 932--940 (2010).
\newblock \doi{10.1016/j.jnnfm.2010.01.016}

\bibitem{GS11}
V.~Garz\'o, A.~Santos, Hydrodynamics of inelastic {Maxwell} models.
\newblock Math. Model. Nat. Phenom. \textbf{6}(4), 37--76 (2011).
\newblock \doi{10.1051/mmnp/20116403}

\bibitem{GT12}
V.~Garz\'o, E.~Trizac, Impurity in a sheared inelastic {M}axwell gas.
\newblock Phys. Rev. E \textbf{85}, 011,302 (2012).
\newblock \doi{10.1103/PhysRevE.85.011302}

\bibitem{GT15}
V.~Garz\'o, E.~Trizac, Generalized transport coefficients for inelastic
  {M}axwell mixtures under shear flow.
\newblock Phys. Rev. E \textbf{92}, 052,202 (2015).
\newblock \doi{10.1103/PhysRevE.92.052202}

\bibitem{GT16}
V.~Garz\'o, E.~Trizac, Tracer diffusion coefficients in a sheared inelastic
  {M}axwell gas.
\newblock J. Stat. Mech. \textbf{073206} (2016).
\newblock \doi{10.1088/1742-5468/2016/07/073206}

\bibitem{GG19}
R.~{G\'omez Gonz\'alez}, V.~Garz\'o, Simple shear flow in granular
  suspensiones: Inelastic {M}axwell models and {BGK}-type kinetic model.
\newblock J. Stat. Mech. \textbf{013206} (2019).
\newblock \doi{10.1088/1742-5468/aaf719}

\bibitem{RS23}
C.~{S\'anchez Romero}, V.~Garz\'o, High-degree collisional moments of inelastic
  {M}axwell mixtures---{A}pplication to the homogeneous cooling and uniform
  shear flow states.
\newblock Entropy \textbf{25}, 222 (2023).
\newblock \doi{10.3390/e25020222}

\bibitem{KSSAOB05}
K.~Kohlstedt, A.~Snezhko, M.V. Sapozhnikov, I.S. Aranson, J.S. Olafsen,
  E.~Ben-Naim, Velocity distributions of granular gases with drag and with
  long-range interactions.
\newblock Phys. Rev. Lett. \textbf{95}, 068,001 (2005).
\newblock \doi{10.1103/PhysRevLett.95.068001}

\bibitem{KS22}
G.M. Kremer, A.~Santos, Granular gas of inelastic and rough {M}axwell
  particles.
\newblock J. Stat. Phys. \textbf{189}, 23 (2022).
\newblock \doi{10.1007/s10955-022-02984-6}

\bibitem{B97}
M.~Babic, Average balance equations for granular materials.
\newblock Intl. J. Eng. Sci. \textbf{35}, 523--548 (1997).
\newblock \doi{10.1016/S0020-7225(96)00094-8}

\bibitem{MHN02}
N.~Mitarai, H.~Hayakawa, H.~Nakanishi, Collisional granular flow as a
  micropolar fluid.
\newblock Phys. Rev. Lett. \textbf{88}, {174}{301} (2002).
\newblock \doi{10.1103/PhysRevLett.88.174301}

\bibitem{BR04}
J.J. Brey, M.J. Ruiz-Montero, Simulation study of the {Green-Kubo} relations
  for dilute granular gases.
\newblock Phys. Rev. E \textbf{70}, 051,301 (2004).
\newblock \doi{10.1103/PhysRevE.70.051301}

\bibitem{NBSG07}
S.H. Noskowicz, O.~Bar-Lev, D.~Serero, I.~Goldhirsch, Computer-aided kinetic
  theory and granular gases.
\newblock EPL \textbf{79}, {60}{001} (2007).
\newblock \doi{10.1209/0295-5075/79/60001}

\bibitem{E66}
A.C. Eringen, Theory of micropolar fluids.
\newblock J. Math. Mech. \textbf{16}, 1--18 (1966).
\newblock \urlprefix\url{https://www.jstor.org/stable/24901466}

\bibitem{L99b}
G.~{\L}ukaszewicz, \emph{Micropolar Fluids. Theory and Applications} (Springer
  Science+Business Media, New York, 1999)

\bibitem{MSD66}
B.J. McCoy, S.I. Sandler, J.S. Dahler, Transport properties of polyatomic
  fluids. {IV}. {T}he kinetic theory of a dense gas of perfectly rough spheres.
\newblock J. Chem. Phys. \textbf{45}(10), 3485--3512 (1966).
\newblock \doi{10.1063/1.1727365}

\bibitem{DT75}
J.S. Dahler, M.~Theodosopulu, in \emph{Non-simple liquids}, \emph{Advances in
  Chemical Physics}, vol. XXXI, ed. by I.~Prigogine, S.A. Rice (Wiley, New
  York, 1975), pp. 155--229

\bibitem{GK91}
D.C. Gaio, G.M. Kremer, Kinetic theory for polyatomic dense gases of rough
  spherical molecules.
\newblock J. Non-Equilib. Thermodyn. \textbf{16}, 357--379 (1991).
\newblock \doi{10.1515/jnet.1991.16.4.357}

\bibitem{K24}
G.M. Kremer.
\newblock Moderately dense granular gas of inelastic rough spheres.
\newblock To be submitted (2024)

\bibitem{P22}
F.B. Pidduck, The kinetic theory of a special type of rigid molecule.
\newblock Proc. R. Soc. Lond. A \textbf{101}, 101--112 (1922).
\newblock \doi{10.1098/rspa.1922.0028}

\bibitem{CC70}
S.~Chapman, T.G. Cowling, \emph{The Mathematical Theory of Non-Uniform Gases},
  3rd edn. (Cambridge University Press, Cambridge, UK, 1970)

\bibitem{K10a}
G.M. Kremer, \emph{{An Introduction to the Boltzmann Equation and Transport
  Processes in Gases}} (Springer, Berlin, 2010)

\bibitem{CLD65}
D.W. Condiff, W.~Lu, J.S. Dahler, Transport properties of polyatomic fluids, a
  dilute gas of perfectly rough spheres.
\newblock J. Chem. Phys. \textbf{42}, 3445--3475 (1965).
\newblock \doi{10.1063/1.1695749}

\bibitem{BGM10}
J.J. Brey, M.I. {Garc\'ia de Soria}, P.~Maynar, Breakdown of hydrodynamics in
  the inelastic {Maxwell} model of granular gases.
\newblock Phys. Rev. E \textbf{82}, {021}{303} (2010).
\newblock \doi{10.1103/PhysRevE.82.021303}

\end{thebibliography}

\end{document}